\documentclass[aps,prd,showpacs,preprintnumbers,amsmath,amssymb,nofootinbib,notitlepage,11pt]{revtex4-1}

\usepackage{graphicx}
\usepackage{color}
\usepackage{amsmath,amssymb}
\definecolor{blue}{rgb}{0,0,0.5}
\usepackage[colorlinks,linkcolor=blue,urlcolor=blue,citecolor=blue]{hyperref}
\usepackage{slashed}

\allowdisplaybreaks

\newcommand{\im}{\mathrm{i}}
\newcommand{\ADG}{A_{\Delta \Gamma}}
\newcommand{\Smumu}{S_{\mu\mu}}

\newcommand{\diff}{\mathrm{d}}
\newcommand{\BsmumuHeading}{\texorpdfstring{\boldmath $B_s\to\mu^+\mu^-$}{Bs->mumu}}

\begin{document}

\title{\BsmumuHeading\ as current and future probe of new physics}

\def\cluster{Excellence Cluster Universe, TUM, Boltzmannstr. 2, 85748 Garching, Germany}
\def\cinci{Department of Physics, University of Cincinnati, Cincinnati, Ohio 45221,USA}

\author{Wolfgang Altmannshofer}
\affiliation{\cinci}

\author{Christoph Niehoff}
\affiliation{\cluster}

\author{David M.\ Straub}
\affiliation{\cluster}

\begin{abstract}\noindent
The rare flavour-changing neutral current decay $B_s\to\mu^+\mu^-$ is
among the most important indirect probes
of new physics at the LHC, as it is strongly suppressed in the Standard Model,
very sensitive to new physics effects,
and theoretically exceptionally clean. We present a thorough state-of-the-art
analysis of the constraints on new physics from present and future measurements
of this decay, focusing on scalar operators.
We show model-independently and in concrete new physics models,
namely the MSSM and two leptoquark scenarios, that a future precise measurement of
the mass-eigenstate rate asymmetry in $B_s\to\mu^+\mu^-$ would allow to disentangle
new physics scenarios that would be indistinguishable based on measurements
of the branching ratio alone. We also highlight the complementarity between
$B_s\to\mu^+\mu^-$ and direct searches in both model classes.
Our numerics is based on the open source code \texttt{flavio} and is made publicly
available.
\end{abstract}


\maketitle

\section{Introduction}

The decay $B_s\to\mu^+\mu^-$ is the rarest $b$ hadron decay ever observed.
As flavour-changing neutral current based on the $b\to s\mu^+\mu^-$ transition,
it is loop- and CKM-suppressed in the Standard Model (SM). Moreover, due to
angular momentum conservation, the SM contribution to the amplitude is
proportional to the muon mass squared, leading to a strong helicity suppression.
This makes the decay extraordinarily sensitive to physics beyond the SM
that gives rise to chirality-changing quark flavour violation, lifting the
helicity suppression and potentially leading to sizable new physics (NP)
effects. At the same time, it is theoretically exceptionally clean for a
$b$ hadron decay. In fact, the hadronic matrix element is fully determined in
terms of the $B_s$ decay constant, which nowadays can be determined to
a precision of  two per cent with lattice QCD
(see e.g.\ \cite{McNeile:2011ng,Dowdall:2013tga}).
The short-distance contributions are known to next-to-next-to leading order
in QCD \cite{Buchalla:1998ba,Hermann:2013kca} and next-to-leading order in the electroweak interactions
\cite{Bobeth:2013tba},
resulting in a remaining non-parametric
uncertainty of only 1.5\% on the branching ratio \cite{Bobeth:2013uxa}.
Thanks to the helicity structure, no photon-mediated hadronic FCNC contributions,
that plague many of the semi-leptonic exclusive decays, are allowed.

After years of searches for this decay with stronger and stronger limits
obtained by the Tevatron experiments CDF and D0,
the huge amount of $b$ quarks produced at the LHC lead to the first observation
of the decay by the LHCb and CMS collaborations \cite{CMS:2014xfa}.
Recently, LHCb even presented the first single-experiment observation
\cite{Aaij:2017vad}.
As we will discuss in Sec.~\ref{sec:exp}, in the next decade an improvement
of the experimental precision by a factor of 7 and of the theoretical precision
by a factor of 3 is not unreasonable. This will allow much more stringent
constraints on NP, or has the potential to unveil significant deviations from
the SM. Nevertheless, by just observing the branching ratio, there are
NP scenarios which cannot be probed; for instance, a scenario leading to a
different sign of the amplitude but to a similar magnitude will not affect the
branching ratio.
This is why the existence of an additional observable, the mass-eigenstate
rate asymmetry $A_{\Delta\Gamma}$ accessible from the untagged time-dependent decay rate,
with complementary dependence on NP contributions, is crucial \cite{DeBruyn:2012wk}.
A main point of this work is to demonstrate the impact of a future measurement
of $A_{\Delta\Gamma}$ on the parameter space of specific NP models,
complementary to the branching ratio as well as to direct searches.

The strong sensitivity of $B_s\to\mu^+\mu^-$ to NP stems in particular from the
fact that scalar operators of the form $(\bar s_L b_R)(\bar\mu_R \mu_L)$ or
$(\bar s_R b_L)(\bar\mu_L \mu_R)$ can lift the helicity suppression present
in the SM, and $B_s\to\mu^+\mu^-$ is the \textit{only} process sensitive to
these operators in a significant way.
This is why we will focus on NP scenarios that generate sizable contributions
to these operators.\footnote{The $B_s\to\mu^+\mu^-$ decay can also give 
important constraints in models {\it without} scalar operators, see e.g.~\cite{Buras:2012dp,Buras:2012jb,Guadagnoli:2013mru,Straub:2013zca}.} 
Such contributions typically arise in one of two ways.
Either from the exchange of
a coloured boson coupling to a lepton-quark current -- a leptoquark; 
or from the exchange of a new, uncoloured scalar particle -- i.e.\ a heavy
Higgs boson -- with tree-level
or loop-induced flavour-changing couplings to quarks.
In supersymmetric (SUSY) extensions of the SM, heavy Higgs exchange is 
particularly interesting as the corresponding scalar amplitudes can be enhanced by $\tan^3\beta$~\cite{Hamzaoui:1998nu,Choudhury:1998ze,Babu:1999hn}.
Also extended Higgs sectors without SUSY can lead to sizable scalar amplitudes
that are enhanced by $\tan^2\beta$~\cite{Logan:2000iv,Buras:2013rqa,Li:2014fea,Buras:2014zga}. 

The rest of this article is organized as follows.
In Sec.~\ref{sec:obs}, we briefly review the observables accessible
in the $B_s\to\mu^+\mu^-$ decay.
In Sec.~\ref{sec:exp}, we discuss the present status of experimental
measurements and the prospects for experimental and theoretical improvements.
These are then used in Sec.~\ref{sec:modelindependent} for a model-independent
analysis of NP parametrized as Wilson coefficients of dimension-six operators,
both from present data and from hypothetical future measurements.
In Secs.~\ref{sec:susy} and \ref{sec:lq}, we analyze in detail the phenomenology
of the $B_s\to\mu^+\mu^-$ decay in the minimal supersymmetric standard model (MSSM),
as an example of a model with scalar operators generated by heavy Higgs exchange,
as well as in two leptoquark models.
In both scenarios, we
highlight the complementarity between the branching ratio and $A_{\Delta\Gamma}$
as well as between $B_s\to\mu^+\mu^-$ and direct searches.
Sec.~\ref{sec:conclusions} contains our conclusions.

\section{Observables in \BsmumuHeading}\label{sec:obs}
In this section let us briefly review the observables accessible in the $B_s\to\mu^+\mu^-$ decay.
For a more detailed discussion we refer e.g. to \cite{Fleischer:2008uj,Buras:2013uqa}.
Since one cannot expect to measure the helicity of the muons in the foreseeable future,
one considers the helicity-averaged decay rate \cite{Buras:2013uqa},
\begin{align}
 \Gamma(B_s(t) \to \mu^+ \mu^-) &= \Gamma(B_s(t) \to \mu^+_\text{L} \mu^-_\text{L})+\Gamma(B_s(t) \to \mu^+_\text{R} \mu^-_\text{R})\\
 &\propto \left[ \cosh\!\left( \frac{y_s t}{\tau_{B_s}} \right) + \Smumu \, \sin(\Delta M t) + A_{\Delta \Gamma} \sinh\!\left(\frac{y_s t}{\tau_{B_s}} \right) \right] \times e^{-t / \tau_{B_s}}\,. \nonumber
\end{align}
Depending on whether the initial flavour of the $B_s$ meson can be determined
(tagged), one can then measure the \textit{untagged} (CP-averaged)
time-dependent rate
\begin{equation}
\Gamma(B_s(t) \to \mu^+ \mu^-) + \Gamma(\bar{B}_s(t) \to \mu^+ \mu^-)
\propto
\left[ \cosh\!\left( \frac{y_s t}{\tau_{B_s}} \right) + A_{\Delta \Gamma} \sinh\!\left(\frac{y_s t}{\tau_{B_s}} \right) \right] \times e^{-t / \tau_{B_s}}\,,
\end{equation}
or the CP asymmetry in the decay rate,
\begin{equation}
\frac{\Gamma(B_s(t) \to \mu^+ \mu^-) - \Gamma(\bar{B}_s(t) \to \mu^+ \mu^-)}{\Gamma(B_s(t) \to \mu^+ \mu^-) + \Gamma(\bar{B}_s(t) \to \mu^+ \mu^-)} = \frac{\Smumu \sin( \Delta M t)}{\cosh\!\left( \frac{y_s t}{\tau_{B_s}} \right) + A_{\Delta \Gamma} \sinh\!\left(\frac{y_s t}{\tau_{B_s}}\right)}.
\end{equation}
Since the width difference in the $B_s$ system is sizable \cite{Amhis:2016xyh},
\begin{equation}
 y_s = \frac{\Delta \Gamma_s}{2 \Gamma_s} = 0.065 \pm 0.005,
\end{equation}
the decay is characterized by three observables:
\begin{itemize}
 \item The mass-eigenstate rate asymmetry
       \begin{equation}
        \ADG = \frac{\Gamma(B_s^\text{H} \to \mu^+\mu^-) - \Gamma(B_s^\text{L} \to \mu^+\mu^-)}{\Gamma(B_s^\text{H} \to \mu^+\mu^-) + \Gamma(B_s^\text{L} \to \mu^+\mu^-)},
       \end{equation}
       where $B_s^\text{H}$ and $B_s^\text{L}$ denote the heavy and light mass eigenstates of the $B_s$ system.
       As seen above, this asymmetry can be extracted from the untagged rate
       and it will be of primary interested in this work.
       For poor statistics, an experimental extraction of this observable is easier in terms of an \emph{effective lifetime} \cite{DeBruyn:2012wk},
       \begin{equation} \label{eq:taumumu}
        \tau_{\mu\mu} = \frac{\int_0^\infty \diff t \,\, t \, \left< \Gamma(B_s(t) \to \mu^+ \mu^-) \right>}{\int_0^\infty \diff t  \, \left< \Gamma(B_s(t) \to \mu^+ \mu^-) \right>},
       \end{equation}
       where $\left< \Gamma(B_s(t) \to \mu^+ \mu^-) \right> = \Gamma(B_s(t) \to \mu^+ \mu^-) + \Gamma(\bar B_s(t) \to \mu^+ \mu^-)$ is the untagged rate.
       The effective lifetime is connected to $\ADG$ via
       \begin{equation}
        \ADG = \frac{1}{y_s} \frac{(1-y_s^2)\tau_{\mu\mu} - (1+y_s^2) \tau_{B_s}}{2 \tau_{B_s} - (1-y_s^2) \tau_{\mu\mu}}.
       \end{equation}
       In the SM these observables take the values
       \begin{equation}
        \ADG^\text{SM} = +1, \qquad\qquad \tau_{\mu\mu}^\text{SM} = \frac{1}{1-y_s} \tau_{B_s} = ( 1.615 \pm 0.010) \, \text{ps}.
       \end{equation}
       In general, $\ADG$ can only take values between $-1$ and $+1$.
 \item The time-integrated (and CP-averaged) branching ratio $\overline{\text{BR}}(B_s \to \mu^+ \mu^-)$.
       It is related to the ``prompt'' branching ratio, i.e.\ the branching ratio in the absence of $B_s$-$\bar B_s$ mixing,
       as~\cite{DeBruyn:2012wk}
       \begin{align} \label{eq:BRprompt}
        \overline{\text{BR}}(B_s \to \mu^+ \mu^-) &= \frac{1}{2} \int \limits_0^\infty \diff t \, \left< \Gamma(B_s(t) \to \mu^+ \mu^-) \right> \nonumber \\
						  &= \frac{1 + \ADG y_s}{1-y_s^2} \,\, \text{BR}(B_s \to \mu^+ \mu^-)_\text{prompt}\,.
       \end{align}
 \item The mixing-induced CP asymmetry $\Smumu$. The measurement of $\Smumu$ requires
 flavour tagging, therefore large amounts of data are needed to overcome small tagging efficiencies
 at LHC. In the SM one has $\Smumu^\text{SM} = 0$. We will not consider $\Smumu$ in the following.
\end{itemize}

\section{Experimental status and prospects}\label{sec:exp}

First evidence for $B_s\to\mu^+\mu^-$ has been found by the LHCb~\cite{Aaij:2013aka}
and CMS~\cite{Chatrchyan:2013bka} experiments individually, who subsequently
combined their measurements of $B_s\to\mu^+\mu^-$ and their searches for
$B^0\to\mu^+\mu^-$ from LHC Run~1 data \cite{CMS:2014xfa}.
Recently, the LHCb collaboration presented their measurement including
Run~2 data \cite{Aaij:2017vad}, representing the first single-experiment observation
of the decay. For $B_s\to\mu^+\mu^-$, the two experiments find\footnote{%
We note that the central value of the CMS Run~1 measurement has been changed
in \cite{CMS:2014xfa} from $3.0\times 10^{-9}$ to $2.8\times 10^{-9}$.
We cannot use this updated result though as the two-dimensional likelihood is
not given. Since the shift is much smaller than the uncertainty, neglecting it
seems well justified. We thank Marc-Olivier Bettler and Patrick Koppenburg for
bringing this point to our attention.
}
\begin{align}
\label{eq:expCMS}
 \overline{\text{BR}}(B_s \to \mu^+\mu^-)_\text{CMS} &= \left( 3.0 ^{+1.0}_{-0.9}\right) \times 10^{-9} ,\\
\label{eq:expLHCb}
 \overline{\text{BR}}(B_s \to \mu^+\mu^-)_\text{LHCb 2017} &= \left( 3.0 \pm 0.6 ^{+0.3}_{-0.2}\right) \times 10^{-9} .
\end{align}
LHCb also performed a first measurement of the effective lifetime, albeit still
with sizable uncertainties that do not allow significant constraints on $A_{\Delta\Gamma}$ yet.
Searches for $B_s\to\mu^+\mu^-$ have also been performed by the CDF~\cite{Aaltonen:2013as},
D0~\cite{Abazov:2013wjb}, and
ATLAS~\cite{Aaboud:2016ire} experiments, the latter being the most sensitive.
For the branching ratio, ATLAS finds from the one-dimensional profile likelihood
\begin{equation}
\overline{\text{BR}}(B_s \to \mu^+\mu^-)_\text{ATLAS} = (0.9^{+1.1}_{-0.8}) \times 10^{-9} ~.
\end{equation}

\begin{figure}[tbp]
\centering
\includegraphics[width=0.9\textwidth]{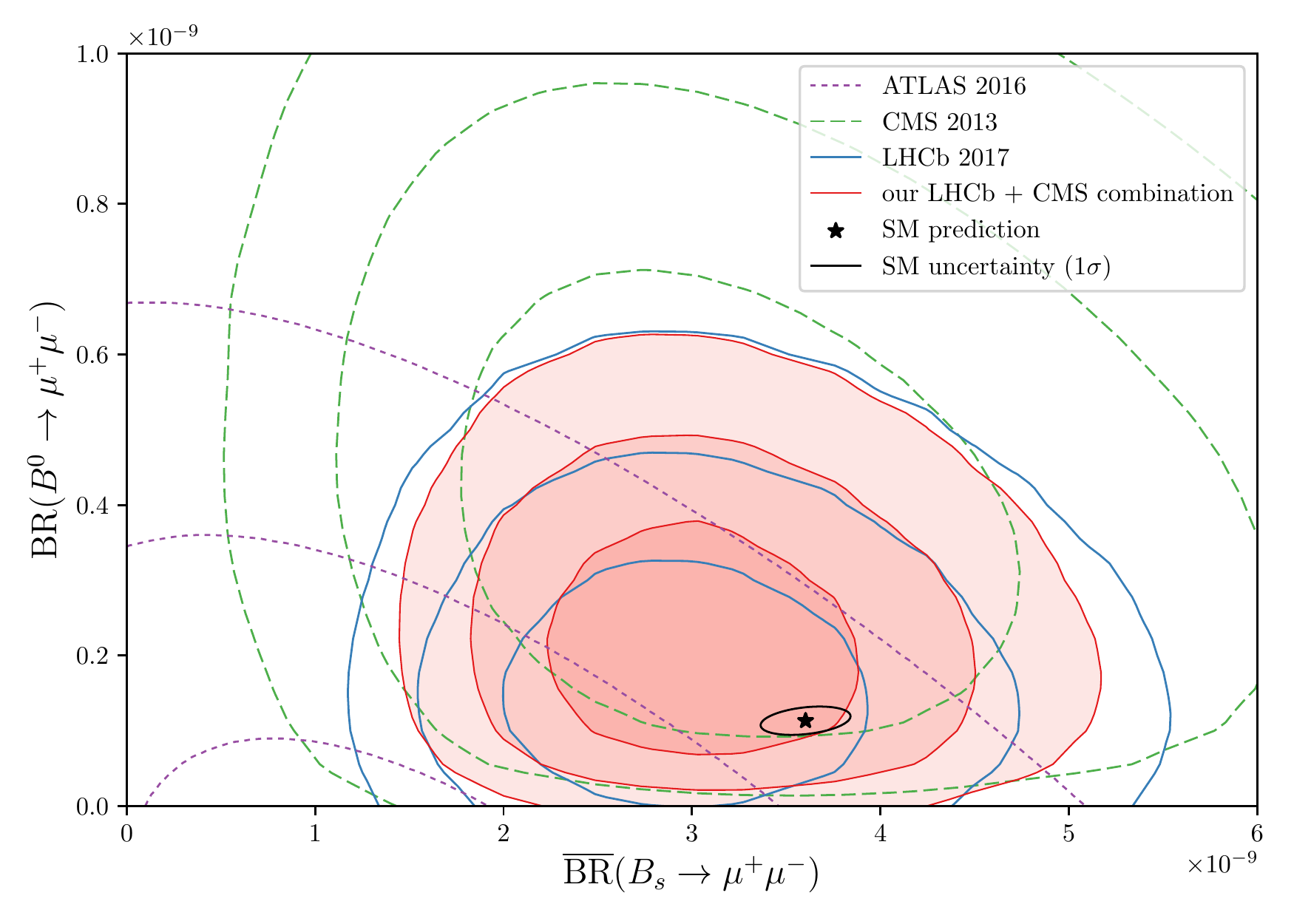}
\caption{Lines of constant likelihood at 1, 2, and $3\sigma$
($-2\,\Delta\ln L\approx2.30,6.18,11.83$)
for the three measurements at LHC experiments as well as for our
``naive'' combination of the Run~1 CMS measurement with the Run~1+2 LHCb
measurement used in our numerics. Also shown are the SM predictions with their correlated $1\sigma$ uncertainties.}
\label{fig:exp}
\end{figure}

In our numerical analysis, we use a combination
of the two measurements that have found evidence for the decay, namely CMS and
LHCb. Since $B_s\to\mu^+\mu^-$ and $B^0\to\mu^+\mu^-$ are correlated, it would
not be consistent to simply average the two numbers in
\eqref{eq:expCMS} and \eqref{eq:expLHCb}, but the two-dimensional likelihoods
have to be combined. While we cannot perform a full combination taking into
account correlations between the experiments, we expect a ``naive'' combination
to be sufficient, given statistical uncertainties are still dominant.
In Fig.~\ref{fig:exp}, we show the two-dimensional likelihood contours in the
plane of $\overline{\text{BR}}(B_s \to \mu^+\mu^-)$ vs. $\text{BR}(B^0 \to \mu^+\mu^-)$
for our combination and the individual measurements, showing also ATLAS
for comparison.

Profiling over $\text{BR}(B^0 \to \mu^+\mu^-)$, we find the one-dimensional $1\sigma$
interval\footnote{%
Fixing $\text{BR}(B^0 \to \mu^+\mu^-)$ to its SM value instead, we find a negligibly
different interval.}
\begin{equation}
\overline{\text{BR}}(B_s \to \mu^+\mu^-)_\text{CMS + LHCb 2017} = (3.00^{+0.55}_{-0.54}) \times 10^{-9} .
\label{eq:ourav}
\end{equation}
This combination is in reasonable agreement with the SM prediction for the branching ratio,
\begin{equation} \label{eq:BRSM}
 \overline{\text{BR}}(B_s \to \mu^+\mu^-)_\text{SM} = \left( 3.60 \pm 0.18 \right) \times 10^{-9} ,
\end{equation}
obtained with \texttt{flavio}~v0.20.4 \cite{flavio} using the calculation of \cite{Bobeth:2013uxa} with
updated hadronic input listed in Tab.~\ref{tab:input}.

\begin{table}
\centering
\renewcommand{\arraystretch}{1.4}
\renewcommand\tabcolsep{12pt}
\begin{tabular}{ccc}
\hline\hline
Parameter & Value & Ref. \\
\hline
$\tau_{B_s}$ & $1.520(4)~\text{ps}$ & \cite{Olive:2016xmw} \\
$\Delta\Gamma_s/\Gamma_s$ & $0.129(9)$& \cite{Amhis:2016xyh}\\
$f_{B_s}$ & $228.4(3.7)~\text{MeV}$ &  \cite{Aoki:2016frl}\\
$|V_{cb}|$ & $4.221(78)\times10^{-2}$  & \cite{Alberti:2014yda}\\
\hline\hline
\end{tabular}
\caption{Numerical inputs for the SM calculation of $\overline{\text{BR}}(B_s \to \mu^+\mu^-)$
using \texttt{flavio}~v0.20.4.}
\label{tab:input}
\end{table}

To arrive at the average \eqref{eq:ourav}, we interpolated between the lines of constant
likelihood in the experimental plots.
We encourage the experimental collaborations to
publish their two-dimensional likelihoods in numerical form in the future.
We have implemented the interpolated numerical two-dimensional likelihoods of the
CMS and LHCb measurements as default constraints on
$\overline{\text{BR}}(B_s \to \mu^+\mu^-)$ and $\text{BR}(B^0 \to \mu^+\mu^-)$
in the public \texttt{flavio} version~0.20.4.

\medskip

The experimental uncertainty on the $B_s \to \mu^+\mu^-$ branching ratio is presently
dominated by statistics and can improve substantially with more data. The LHCb
collaboration expects to measure it with a statistical precision of
$0.19 \times 10^{-9}$
for an integrated luminosity of 50~fb$^{-1}$~\cite{LHCb-PUB-2014-040},
corresponding to their expectation for LHC Run~4,
while CMS expects a precision of $0.4 \times 10^{-9}$ for 3000~fb$^{-1}$ \cite{CMS:2015iha},
corresponding to their expectation for Run~5.
An upgrade of LHCb for Run~5 that would allow the experiment to collect an
integrated luminosity of 300~fb$^{-1}$ could reduce the statistical precision,
applying naive scaling, to $0.08 \times 10^{-9}$, corresponding to around
2\% of the SM prediction.

While the expectations for the branching ratio might be affected by overall
systematic uncertainties stemming, for instance, from the $B_s$ production
fraction or normalization modes, such overall uncertainties cancel in the
determination of $A_{\Delta \Gamma}$. On the other hand,
$A_{\Delta \Gamma}$ is suppressed by the small $B_s$-$\bar B_s$
lifetime difference. A measurement of the time-dependent decay rate with
a precision of 5\% (2\%), assuming the SM, thus translates into uncertainties on
$A_{\Delta \Gamma}$ of $0.8 ~(0.3)$.
Optimistically assuming that the systematic uncertainties on the branching ratio
can be sufficiently reduced, we thus arrive at the following future scenarios,
\begin{subequations}
 \begin{eqnarray}
   && \sigma_\text{exp}(B_s \to \mu^+\mu^-) = 0.19 \times 10^{-9}  \,,~~ \sigma_\text{exp}(A_{\Delta \Gamma}) = 0.8 \,, ~~~\text{for 50~fb$^{-1}$ ~\,(``Run 4'')} , \label{eq:run4}\\
   && \sigma_\text{exp}(B_s \to \mu^+\mu^-) = 0.08 \times 10^{-9}  \,,~~ \sigma_\text{exp}(A_{\Delta \Gamma}) = 0.3 \,, ~~~\text{for 300~fb$^{-1}$ (``Run 5'')} . \label{eq:run5}
 \end{eqnarray} \label{eq:UncertaintyProjections}
\end{subequations}
On the theory side, the current relative uncertainty is dominated by the
uncertainties on the CKM element $V_{cb}$ and the $B_s$ decay constant $f_{B_s}$.
On a time scale of ten years necessary for the projected experimental improvements,
a lattice computation of $f_{B_s}$ to a precision of 1~MeV seems realistic
\cite{Charles:2013aka,latticewhitepaper}. Similarly a lattice computation of the $B\to D$ form factors to a
precision of 0.7\% might be possible, which would allow to extract $V_{cb}$ from $B\to D\ell\nu$
decays with this precision, given a sufficiently precise measurement at Belle-II.
Assuming that also the remaining non-parametric uncertainties \cite{Bobeth:2013uxa}
can be reduced
by a factor of two to three,
a theoretical precision of 1.6\% on the SM branching ratio could
be reached, competitive with experiment. The theory uncertainty on
$A_{\Delta \Gamma}$ is instead negligible compared to the experimental
expectations. We thus take
\begin{eqnarray}
 && \sigma_\text{th}(B_s \to \mu^+\mu^-) = 0.06 \times 10^{-9}  \,,~~ \sigma_\text{th}(A_{\Delta \Gamma}) \approx 0
 \label{eq:thproj}
\end{eqnarray}
for both future scenarios.

\section{Model-independent analysis}\label{sec:modelindependent}

Before analyzing specific NP models, in this section we perform a
model-independent analysis of new physics effects in $B_s\to\mu^+\mu^-$.
The relevant effective Hamiltonian for the decay valid in any extension of
the SM without new particles near or below the $b$ quark mass scale reads
\begin{equation}
\mathcal{H}_\text{eff} = - \frac{4G_F}{\sqrt{2} \pi} V_{ts}^* V_{tb} \frac{e^2}{16\pi^2}
\sum_{i\in[10,S,P]}\left[ C_i O_i + C_i' O_i' + \text{h.c.} \right],
\label{eq:Heff}
\end{equation}
with
\begin{align}
O_{10}^{(\prime)}  &=     (\bar{s}_{L(R)} \gamma_\rho b_{L(R)})( \bar{\mu} \gamma^\rho \mu), &
O_S^{(\prime)}     &= m_b (\bar{s}_{L(R)} b_{R(L)})(\bar{\mu} \mu), &
O_P^{(\prime)}     &= m_b (\bar{s}_{L(R)} b_{R(L)})(\bar{\mu} \gamma_5 \mu),
\end{align}
where only $C_{10}$ is non-vanishing in the SM.
The normalization of $O_{S,P}^{(\prime)}$ is chosen to make their Wilson
coefficients renormalization group invariant. We thus do not need to specify
the renormalization scale for the Wilson coefficients in the following, but
note that $C_{S,P}^{(\prime)}$ have dimensions of inverse mass in this
convention.

The prompt branching ratio (as used in (\ref{eq:BRprompt})) and $A_{\Delta\Gamma}$ are given in terms of the Wilson coefficients as
\begin{align}
 \text{BR}(B_s \to \mu^+ \mu^-)_\text{prompt} &= \frac{G_F^2 \alpha^2}{16 \pi^3} \left| V_{ts} V_{tb}^* \right|^2 f_{B_s}^2 \tau_{B_s} m_{B_s} m_\mu^2 \sqrt{1 - 4 \frac{m_\mu^2}{m_{B_s}^2}} \left| C_{10}^\text{SM} \right|^2 \left( \left| P \right|^2 + \left| S \right|^2 \right), \\
 A_{\Delta\Gamma}    &= \frac{\left| P \right|^2 \cos(2 \phi_P - \phi_s^\text{NP}) - \left| S \right|^2 \cos(2 \phi_S - \phi_s^\text{NP})}{\left| P \right|^2 + \left| S \right|^2},
\end{align}
where the Wilson coefficients appear through the combinations
\begin{align}
P &= \frac{C_{10} - C'_{10}}{C_{10}^\text{SM}} + \frac{M^2_{B_s}}{2 m_\mu} \frac{m_b}{m_b + m_s} \left( \frac{C_P - C'_P}{C_{10}^\text{SM}} \right), &
S &= \sqrt{1-4\frac{m_\mu^2}{M^2_{B_s}}} \frac{M^2_{B_s}}{2 m_\mu} \frac{m_b}{m_b + m_s} \left( \frac{C_S - C'_S}{C_{10}^\text{SM}} \right),
\end{align}
with $S = \left| S \right| \exp(\im \phi_S)$,  $P = \left| P \right| \exp(\im \phi_P)$,
and $\phi_s^\text{NP}$ is a NP contribution to the $B_s^0-\bar{B}_s^0$ mixing phase.
In the SM one has $S=0$ and $P=1$.

While the effective Hamiltonian \eqref{eq:Heff} is appropriate for any NP
that is heavy compared to the $b$ quark mass scale, the absence of any signal
for NP at LHC typically implies that the NP scale is much larger than the
electroweak scale. If this is the case, one can express NP effects model-independently
in terms of Wilson coefficients of an effective Lagrangian invariant under
the full SM gauge symmetry. At dimension six, the matching of this ``SMEFT''
\cite{Buchmuller:1985jz,Grzadkowski:2010es}
onto the above weak effective Hamiltonian leads to two relations among the
Wilson coefficients \cite{Alonso:2014csa},
\begin{equation} \label{eq:GSMinvariance}
 C_S = - C_P, \qquad C_S' = C_P'.
\end{equation}
In models where the electroweak symmetry is realized non-linearly in the Higgs
sector, these relations can be violated \cite{Cata:2015lta}.
However, in realistic models with
strongly-coupled electroweak symmetry breaking where the Higgs is a pseudo
Goldstone boson, this breaking is suppressed by $v^2/f^2$, where $f$ is the
pseudo Goldstone decay constant. Although formally of order one, this
ratio is usually required to be below the few percent level to comply with
precision constraints in the electroweak and Higgs sectors. Therefore we expect
the relations~\eqref{eq:GSMinvariance} to hold to good accuracy in most
realistic models and we will assume them to hold for the rest of this section.

In models with NP effects in $C_{10}$ or $C_{10}'$, but not in the scalar
operators, $B_s\to\mu^+\mu^-$ plays an important role as it allows to measure
the combination $C_{10}-C_{10}'$ with smaller theoretical uncertainties and
without dependence on possible NP effects in electromagnetic penguins or
semi-leptonic vector operators (usually called $O_9^{(\prime)}$).
Here, we want to focus instead on NP effects in the scalar operators, since
$B_s\to\mu^+\mu^-$ has the unique property of being strongly helicity suppressed
in the SM, but not in the presence of NP in scalar operators.
In semi-leptonic decays, only two observables are sensitive to scalar operators
with muons in principle:
\begin{itemize}
 \item The ``flat term'' $F_H$ in $B\to K\mu^+\mu^-$ is tiny in the SM in the absence
 of scalar operators \cite{Bobeth:2007dw,Becirevic:2012ec,Beaujean:2015gba}.
 However it turns out that if relations \eqref{eq:GSMinvariance}
 is satisfied, the observable is \textit{not} complementary to $B_s\to\mu^+\mu^-$
 and not competitive.\footnote{A model-independent analysis of scalar operators without the relation~(\ref{eq:GSMinvariance}) has been performed recently in \cite{Beaujean:2015gba}.}
 \item The angular observable $S_6^c$ in $B\to K^*\mu^+\mu^-$ \cite{Altmannshofer:2008dz}.
 Again however, the effects are too small to be competitive with $B_s\to\mu^+\mu^-$.
\end{itemize}
Thus we will restrict ourselves to $B_s\to\mu^+\mu^-$.

We now proceed to an analysis of NP effects in the two independent Wilson
coefficients $C_S$ and $C_S'$, assuming $C_{10}^{(\prime)}$ to be SM-like.
For the numerics, we used \texttt{flavio} v0.20.4 \cite{flavio}.

\subsection{Constraints on a single Wilson coefficient}

We start by considering NP effects in a single real Wilson coefficient $C_S = - C_P$.
This applies to all models with heavy NP
(and linearly realized electroweak symmetry in the Higgs sector)
that satisfy the criterion of Minimal Flavour Violation \cite{DAmbrosio:2002vsn}, in particular
the MFV MSSM to be analyzed in Sec.~\ref{sec:susy}.
Within this framework, we have to an excellent approximation $S \simeq 1-P \equiv A$, with a real $A$.
The $B_s \to \mu^+\mu^-$ branching ratio, BR$(B_s \to \mu^+\mu^-)$, and the mass-eigenstate rate asymmetry, $A_{\Delta \Gamma}$ can therefore be written as
\begin{eqnarray} \label{eq:BR_A}
 \frac{\overline{\text{BR}}(B_s \to \mu^+\mu^-)}{\overline{\text{BR}}(B_s \to \mu^+\mu^-)_\text{SM}} &=&  (1 - A)^2 + A^2 - \frac{y_s}{1+y_s} 2 A^2  ~, \\ \label{eq:ADG_A}
 A_{\Delta \Gamma} &=& \frac{(1-A)^2 - A^2}{(1-A)^2 + A^2} ~.
\end{eqnarray}
There are two regions of parameter space that correspond to a SM-like branching ratio. The first one corresponds to a small new physics contribution $A \ll 1$ and has $A_{\Delta \Gamma} \simeq A_{\Delta \Gamma}^\text{SM} = 1$. The second region corresponds to a NP contribution that is comparable to the SM amplitude $A \simeq 1$. This second region of parameter space predicts $A_{\Delta \Gamma} \simeq -1$. While measurements of the branching ratio alone cannot distinguish the two regions, a measurement of $A_{\Delta \Gamma}$ can.

\begin{figure}[tbp]
\centering
\includegraphics[width=0.8\textwidth]{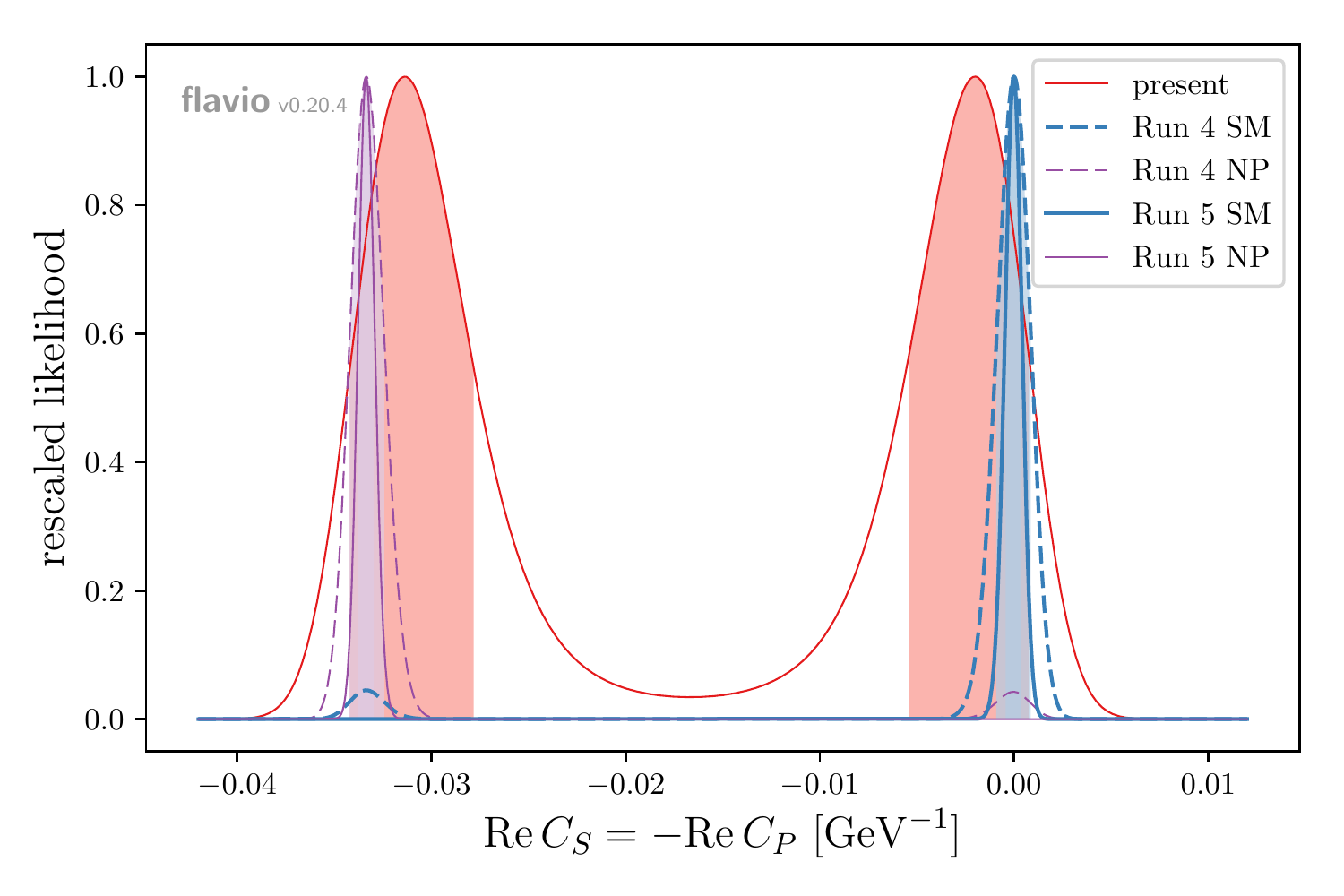}
\caption{Present and future constraints on the real part of the Wilson
coefficient $C_S$, assumed to satisfy the SMEFT relation. The vertical
axis corresponds to the likelihood (containing experimental and theoretical
uncertainties), rescaled to equal maximum likelihood.
The scenarios labeled ``Run 4'' (dashed lines) correspond to eq.~\eqref{eq:run4},
``Run 5'' (solid lines) to eq.~\eqref{eq:run5}.
The ``NP'' scenario (thin lines) predicts the same branching
ratio as in the SM (thick lines), but opposite $A_{\Delta\Gamma}$.
The shaded ranges correspond to $1\sigma$ (highest likelihood
regions containing $68.3\%$ of the integrated likelihood).}
\label{fig:modelindependent:OneWC}
\end{figure}

We demonstrate this by performing the following fits of this Wilson coefficient:
\begin{itemize}
 \item A fit to the present branching ratio measurement,
 \item A fit to a future measurement with the projected uncertainties in
 \eqref{eq:run4} and \eqref{eq:thproj} assuming the experimental central values
 for the branching ratio and $A_{\Delta\Gamma}$ to equal the SM central values
 (``SM scenario'' with $A\ll 1$),
 \item A similar ``future'' fit assuming the measured branching ratio to coincide
 with the SM expectation, but the central value of $A_{\Delta\Gamma}$ to be $-1$
 (``NP scenario'' with $A\approx 1$),
 \item The previous two fits also for the future scenario in \eqref{eq:run5}.
\end{itemize}
For these fits, we have combined the correlated
experimental and theoretical uncertainties into a likelihood function depending
only on the Wilson coefficient, as described in \cite{Altmannshofer:2014rta}
(and implemented in \texttt{flavio} as \texttt{FastFit}).
The result is shown in Fig.~\ref{fig:modelindependent:OneWC}. We observe that
\begin{itemize}
\item the current measurement leaves two solutions with equal likelihood\footnote{%
We do not take into account the 2017 LHCb measurement of the effective lifetime
that distinguishes the two solutions at less than half a standard deviation,
\cite{Aaij:2017vad}.},
\item future measurements of $A_{\Delta\Gamma}$ will be able to completely
exclude one of the two solutions, depending on the sign
of $A_{\Delta\Gamma}$.
\end{itemize}

\subsection{Constraints on a pair of Wilson coefficients}

In the leptoquark models to be studied in Sec.~\ref{sec:lq}, both coefficients
$C_S$ and $C_S'$ can be generated simultaneously. We thus proceed to analyze
simultaneous NP effects in these two coefficients, first assuming them to be real.

\begin{figure}[tbp]
\centering
\includegraphics[width=0.49\textwidth]{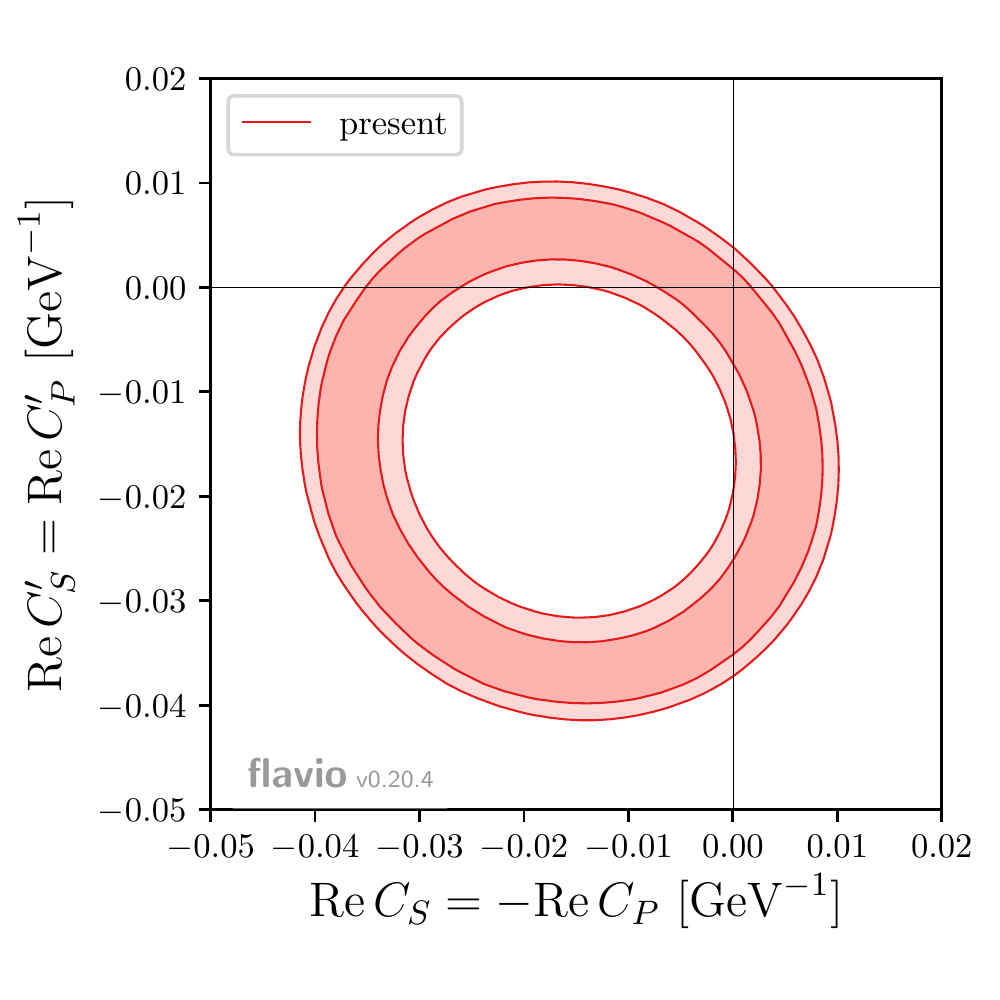}\\
\includegraphics[width=0.49\textwidth]{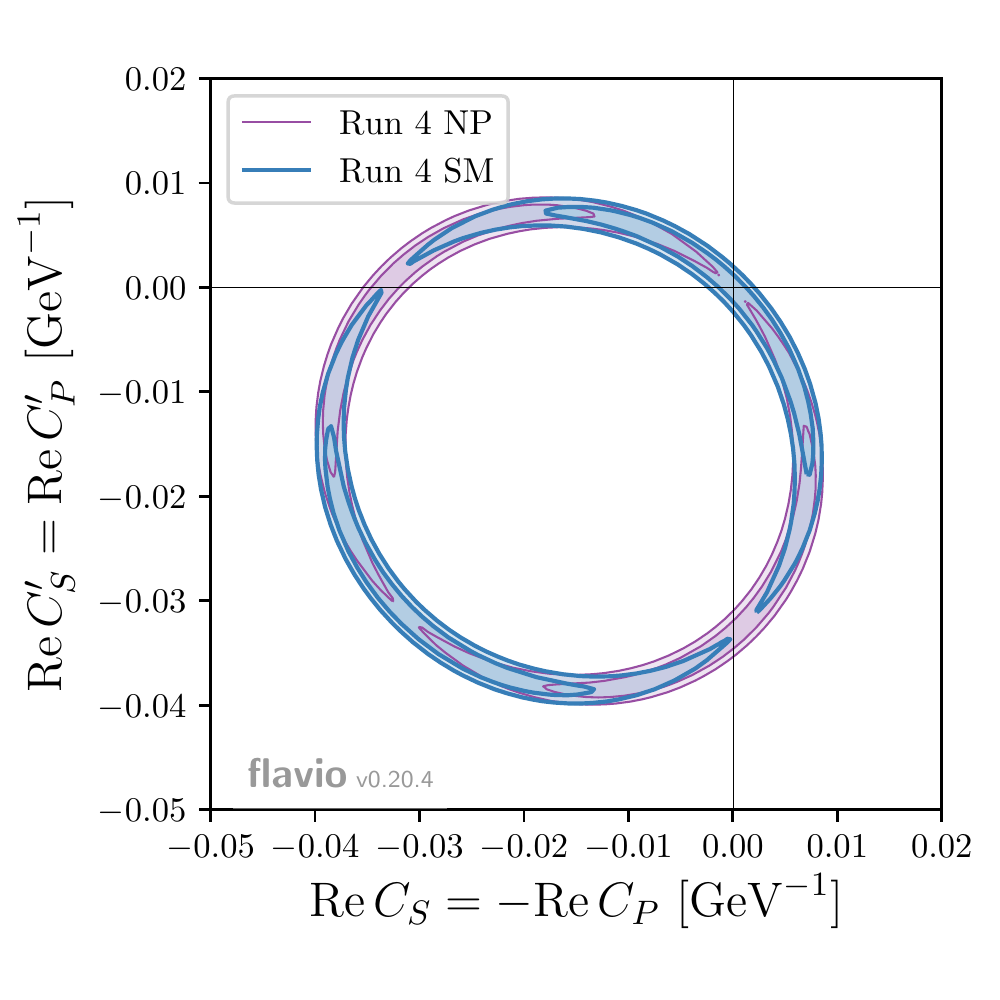} ~
\includegraphics[width=0.49\textwidth]{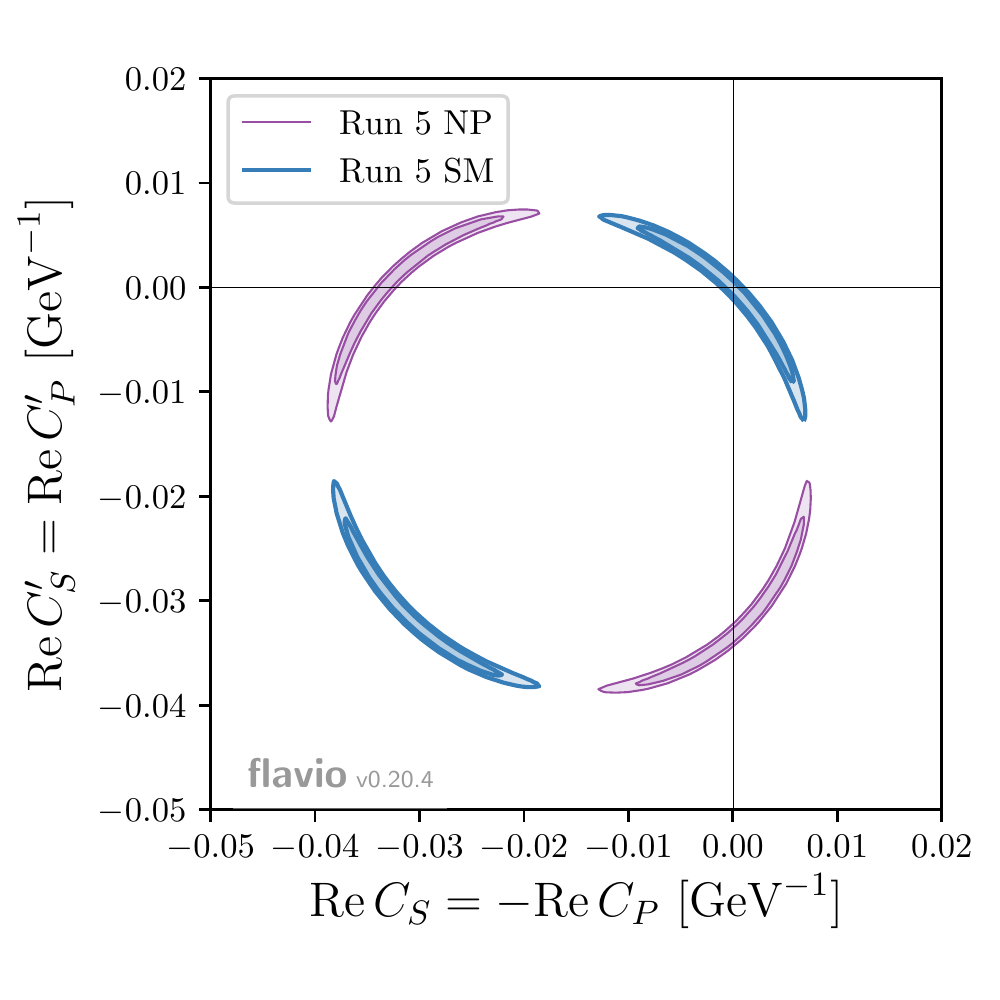}
\caption{Present (top) and future (bottom) constraints on the real parts of the Wilson
coefficients $C_S$ and $C_S'$, assumed to satisfy the SMEFT relation~(\ref{eq:GSMinvariance}),
showing the 1 and $2\sigma$ ($-2\,\Delta\ln L\approx2.30$ and $6.18$) contours.
The NP scenario is the same as in Fig.~\ref{fig:modelindependent:OneWC}.}
\label{fig:modelindependent:TwoWCs}
\end{figure}

Just as for the single Wilson coefficient, we performed five fits: to present
data and to two future scenarios with SM-like or opposite-sign $A_{\Delta\Gamma}$.
The results are shown in Fig.~\ref{fig:modelindependent:TwoWCs}. We make
the following observations.
\begin{itemize}
\item The degeneracy from the present branching ratio measurement leads to a
ring in the $C_S$-$C_S'$ plane.
\item Future measurements of $A_{\Delta \Gamma}$ can break this degeneracy, but a two-fold ambiguity
cannot be resolved by the branching ratio and $A_{\Delta\Gamma}$.
\item The precision of the future scenario of \eqref{eq:run5} (``Run 5'')
is required to obtain disjoint solutions at $2\sigma$.
\end{itemize}

\begin{figure}[tbp]
\centering
\includegraphics[width=0.49\textwidth]{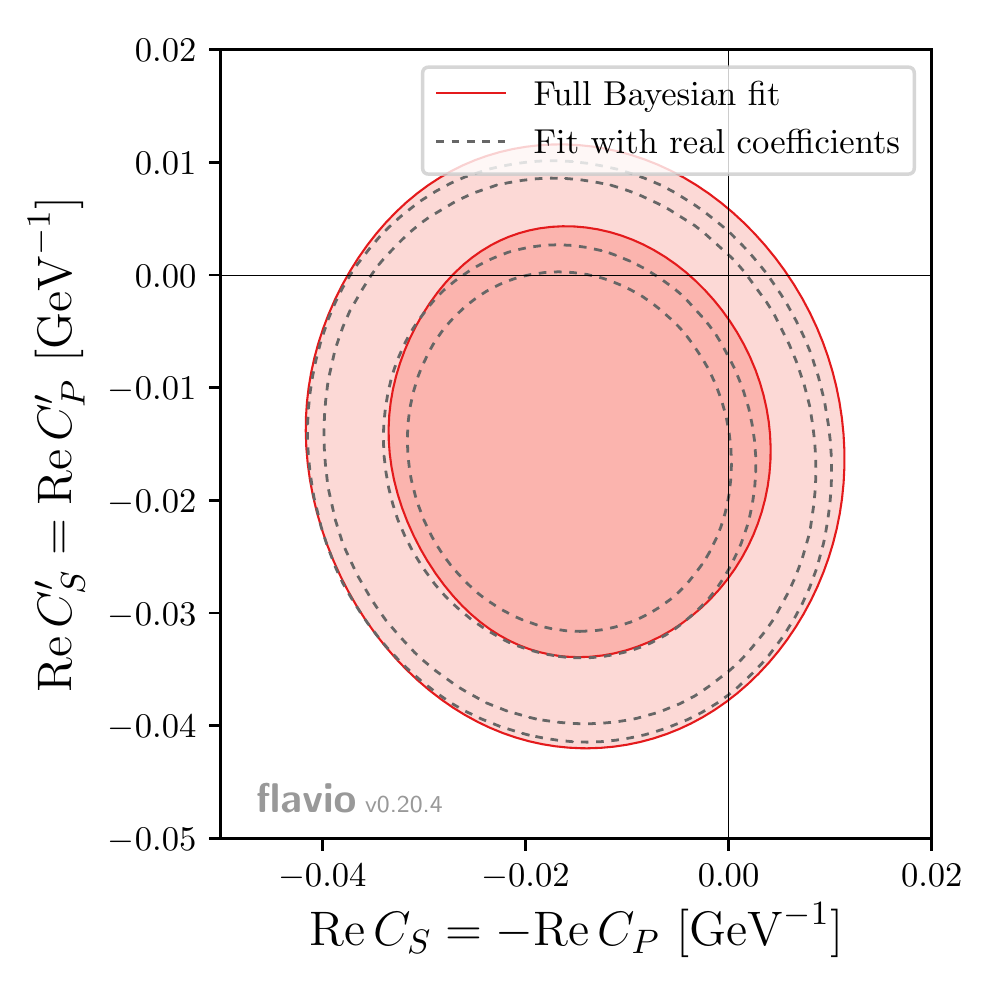}%
\includegraphics[width=0.51\textwidth]{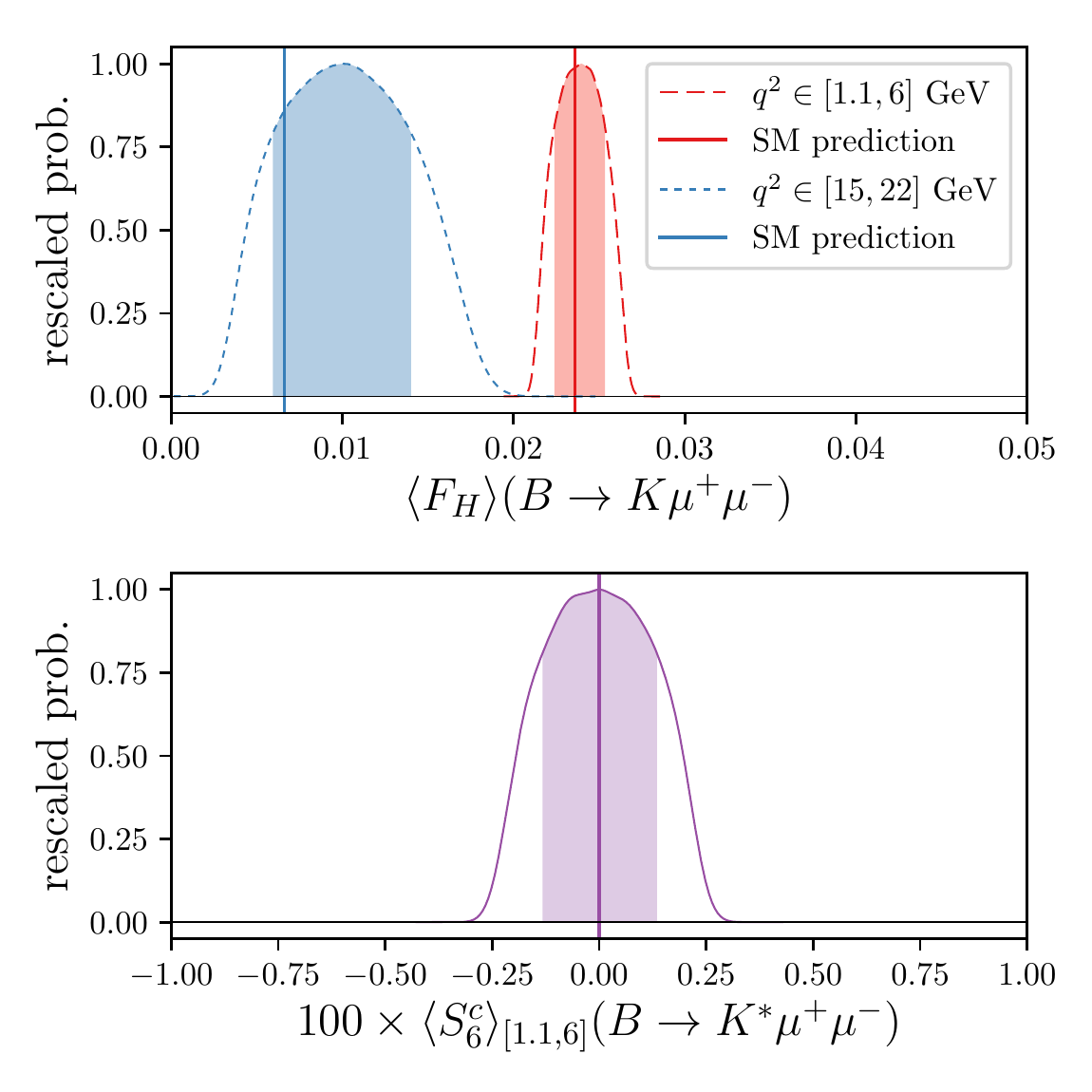}
\caption{Left: marginal posterior distribution from a Bayesian fit with flat priors
for the real and imaginary parts of the Wilson
coefficients $C_S$ and $C_S'$, assumed to satisfy the SMEFT relation~(\ref{eq:GSMinvariance}),
showing the 1 and $2\sigma$ ($68.3\%$ and $95.5\%$ posterior probability) contours.
For comparison, the case of real Wilson coefficients from
Fig.~\ref{fig:modelindependent:TwoWCs} is shown as dashed contours.
Right: posterior prediction from the Bayesian fit for the flat term $F_H$ in
$B\to K\mu^+\mu^-$ at high and low $q^2$ (rescaled to the same
maximum probability), and the angular observable
$S_6^c$ in $B\to K^*\mu^+\mu^-$ at low $q^2$ compared to their
SM predictions.}
\label{fig:modelindependent:Bayes}
\end{figure}

So far, we have assumed the Wilson coefficients to be real. In fact, being
CP-averaged quantities, the branching ratio and $A_{\Delta\Gamma}$ are not
very sensitive to the imaginary parts of $C_{S,P}$ and $C_{S,P}'$,
which do not interfere with the real SM coefficient. Nevertheless, for sizable
complex NP contributions to the scalar Wilson coefficients, the qualitative picture shown
in Fig.~\ref{fig:modelindependent:TwoWCs} changes. To demonstrate this, we have
performed a full Bayesian fit to the current branching ratio data -- using \texttt{flavio} in combination
with \texttt{emcee} \cite{ForemanMackey:2012ig} -- allowing real and imaginary NP contributions to
$C_S$ and $C_S'$ (still satisfying the SMEFT relations \eqref{eq:GSMinvariance}),
with flat priors, and varying also the relevant nuisance parameters such as
$f_{B_s}$ and $y_s$.
The resulting marginal posterior distribution for the real
parts is shown in Fig.~\ref{fig:modelindependent:Bayes} left.
Compared to the case of real NP contributions shown in
Fig.~\ref{fig:modelindependent:TwoWCs} and underlayed as dashed contours here
for comparison, the region within the ``ring'' is now filled, as a too small
branching ratio from the destructive interference between the SM coefficient
and the real parts of the NP coefficients can be compensated by a contribution
from the imaginary parts.

The Bayesian fit can also be used to derive posterior \textit{predictions}
for other observables. In Fig.~\ref{fig:modelindependent:Bayes} right, we show
the posterior predictions for the flat term in $B\to K\mu^+\mu^-$ at high
and low $q^2$,
which is very small in the absence of scalar operators,
as well
as for the angular observable $S_6^c$ in $B\to K^*\mu^+\mu^-$, which
vanishes in the absence of scalar operators.
Obviously, the effects still allowed for a completely general variation
of the real and imaginary parts of $C_S$ and $C_S'$ satisfying the
SMEFT relations \eqref{eq:GSMinvariance} are tiny in both cases, demonstrating
that these observables cannot compete with $B_s\to\mu^+\mu^-$ now or in the
future, as anticipated.

All the plots in this section can be reproduced with \texttt{flavio}~v0.20.4 using
the scripts provided by us in a public repository \cite{plotrepo}.

\section{\BsmumuHeading\ in SUSY Models}\label{sec:susy}

The $B_s \to \mu^+\mu^-$ decay is very well recognized as an important probe of supersymmetric extensions of the SM, as the decay rate can in principle be enhanced by orders of magnitude~\cite{Hamzaoui:1998nu,Choudhury:1998ze,Babu:1999hn}. Therefore, the good agreement between SM prediction and the experimental results for BR$(B_s \to \mu^+\mu^-)$ leads to strong constraints on the parameter space of the minimal supersymmetric standard model. Here we will discuss the impact of the current and expected future $B_s \to \mu^+\mu^-$ measurements in the context of the MSSM with minimal flavour violation, where the only sources of flavour violation are the SM Yukawa couplings. We will make the additional assumption that there are no sources of CP violation beyond the phase in the CKM matrix and that the squark masses are completely flavour blind, i.e. proportional to the unit matrix. In this case, FCNCs are proportional to Higgsino-stop loops. Despite the absence of any new sources of flavour
violation, the $B_s \to \mu^+\mu^-$ decay can provide powerful constraints in this scenario.

In the MSSM with MFV and no new sources of CP violation, the NP contributions can be described by a single, real amplitude $A$ as in eqs.~(\ref{eq:BR_A}) and~(\ref{eq:ADG_A}).
The NP amplitude $A$ is induced by the exchange of heavy scalar and pseudoscalar Higgs bosons, $H$ and $A$. In the decoupling limit, these states are approximately degenerate, $m_H \simeq m_A$. Their combined contribution can be concisely written as~\cite{Altmannshofer:2012ks}
\begin{equation}
 A = \frac{4\pi}{\alpha_2} \frac{m_{B_s}^2}{4 m_A^2} \frac{\epsilon_\text{FC} t_\beta^3}{(1+\epsilon_b t_\beta)(1+\epsilon_0 t_\beta)(1 + \epsilon_\ell t_\beta)} \frac{1}{Y_0} ~,
\end{equation}
where $Y_0 \simeq 0.95$ is a SM loop function. The above amplitude is proportional to the third power of $t_\beta = \tan\beta = v_2/v_1$, the ratio of the two Higgs vacuum expectation values, and decouples with the mass squared of the heavy Higgs bosons, $m_A^2$.
The various $\epsilon$ parameters correspond to loop induced non-holomorphic Higgs couplings. We take into account gluino-squark loops, Wino-squark loops, and Higgsino-stop loops.
General expressions for these contributions can be found in~\cite{Altmannshofer:2012ks}. Here we will work in the limit where all SUSY masses are degenerate. In this simple benchmark case we have
\begin{equation}
 \epsilon_0 = \epsilon_{\tilde g} + \epsilon_{\tilde w} ~,~~ \epsilon_b = \epsilon_{\tilde g} + \epsilon_{\tilde w} + \epsilon_{\tilde h} ~,~~ \epsilon_\ell = \epsilon_{\tilde w} ~,~~ \epsilon_\text{FC} = \epsilon_{\tilde h} ~,
\end{equation}
\begin{equation} \label{eq:epsilons}
 \epsilon_{\tilde g} = \frac{\alpha_s}{4\pi} \frac{4}{3} \times \text{sgn}(\mu m_{\tilde g}) ~,~~
 \epsilon_{\tilde w} = -\frac{\alpha_2}{4\pi} \frac{3}{4} \times \text{sgn}(\mu m_{\tilde w}) ~,~~
 \epsilon_{\tilde h} = \frac{\alpha_2}{4\pi} \frac{m_t^2}{4 m_W^2} \times \frac{A_t}{\mu} ~.
\end{equation}
In the more general case, where SUSY masses are not degenerate, the above expressions are modified by $O(1)$ loop functions that depend on the ratios of SUSY masses.
The signs of the gluino, wino, and higgsino contributions are set by the relative signs of the $\mu$ term and the gluino mass $m_{\tilde g}$, the $\mu$ term and the wino mass $m_{\tilde w}$, and the $\mu$ term and the stop trilinear coupling $A_t$, respectively. We follow the same sign conventions as in~\cite{Altmannshofer:2012ks}.

Note that the above expressions for the $\epsilon$ parameters do not depend directly on the overall SUSY mass scale. A mild dependence on the SUSY scale enters only through normalization group running, as the parameters $\alpha_s$, $\alpha_2$ and $m_t$ in~(\ref{eq:epsilons}) correspond to $\overline{\text{MS}}$ values at the SUSY scale\footnote{In our numerical analysis we use SM 2-loop RGEs~\cite{Machacek:1983tz,Machacek:1983fi,Machacek:1984zw,Luo:2002ey} to run these parameters from the electro-weak scale to the SUSY scale.}. Therefore, bounds from direct searches for SUSY particles, in particular squarks and gluinos, can be easily avoided by decoupling, without significantly affecting the $B_s \to \mu^+\mu^-$ phenomenology. The null-results of existing searches for squarks and gluinos~(see e.g.~\cite{Aad:2015iea,ATLAS:2016kts,ATLAS:2016kjm,ATLAS:2016uzr,CMS:2016inz,CMS:2016xva}) constrain the corresponding particle masses in some cases up to almost $\sim 2$~TeV. Sensitivities will likely increase further
for the
future high luminosity scenarios.
In the following, we will set the SUSY masses to 5~TeV, which should be safely outside the reach even of the high luminosity LHC.

While such heavy SUSY masses require a fine-tuning of the electro-weak scale at levels of below $1\%$, they are easily compatible with a 125~GeV Higgs mass, without having to resort to maximal stop mixing.
Indeed, for stop masses of 5~TeV and trilinear couplings of the same order, the Higgs mass is in the ballpark of 125~GeV over a broad range of $\tan\beta$ values.

In the following we will concentrate on a specific benchmark scenario, to illustrate the complementarity of the branching ratio and the mass-eigenstate rate asymmetry in probing the MSSM parameter space.
We choose $\text{sgn}(\mu m_{\tilde g}) = \text{sgn}(\mu m_{\tilde w}) = - \text{sgn}(\mu A_t) = -1$, corresponding to a positive SUSY amplitude $A$, and fix the absolute value of the stop trilinear coupling $A_t$ by demanding that $m_h = 125$~GeV. Using the \verb|SusyHD| code~\cite{Vega:2015fna}, we find that $|A_t|$ is almost independent of the heavy Higgs mass $m_A$. It lies in the range $|A_t|\in (6.2,7.5)$~TeV for $\tan\beta$ between 10 and 60, and grows to $|A_t| \simeq 10.5$~TeV for $\tan\beta = 80$.
This leads to the following approximate values for the $\varepsilon$ factors that enter the $B_s \to \mu^+\mu^-$ amplitude
\begin{equation}
-0.0039 \lesssim \epsilon_b \lesssim -0.0023 ~,~~ \epsilon_0 \simeq -0.0062 ~,~~ \epsilon_\ell \simeq +0.0019 ~,~~ +0.0023 \lesssim \epsilon_\text{FC} \lesssim +0.0039 ~.
\end{equation}
The ranges for $\epsilon_b$ and $\epsilon_\text{FC}$ are due to the variation of $A_t$ for $10 < \tan\beta < 80$.

\begin{figure}[tb]
\centering
\includegraphics[width=0.46\textwidth]{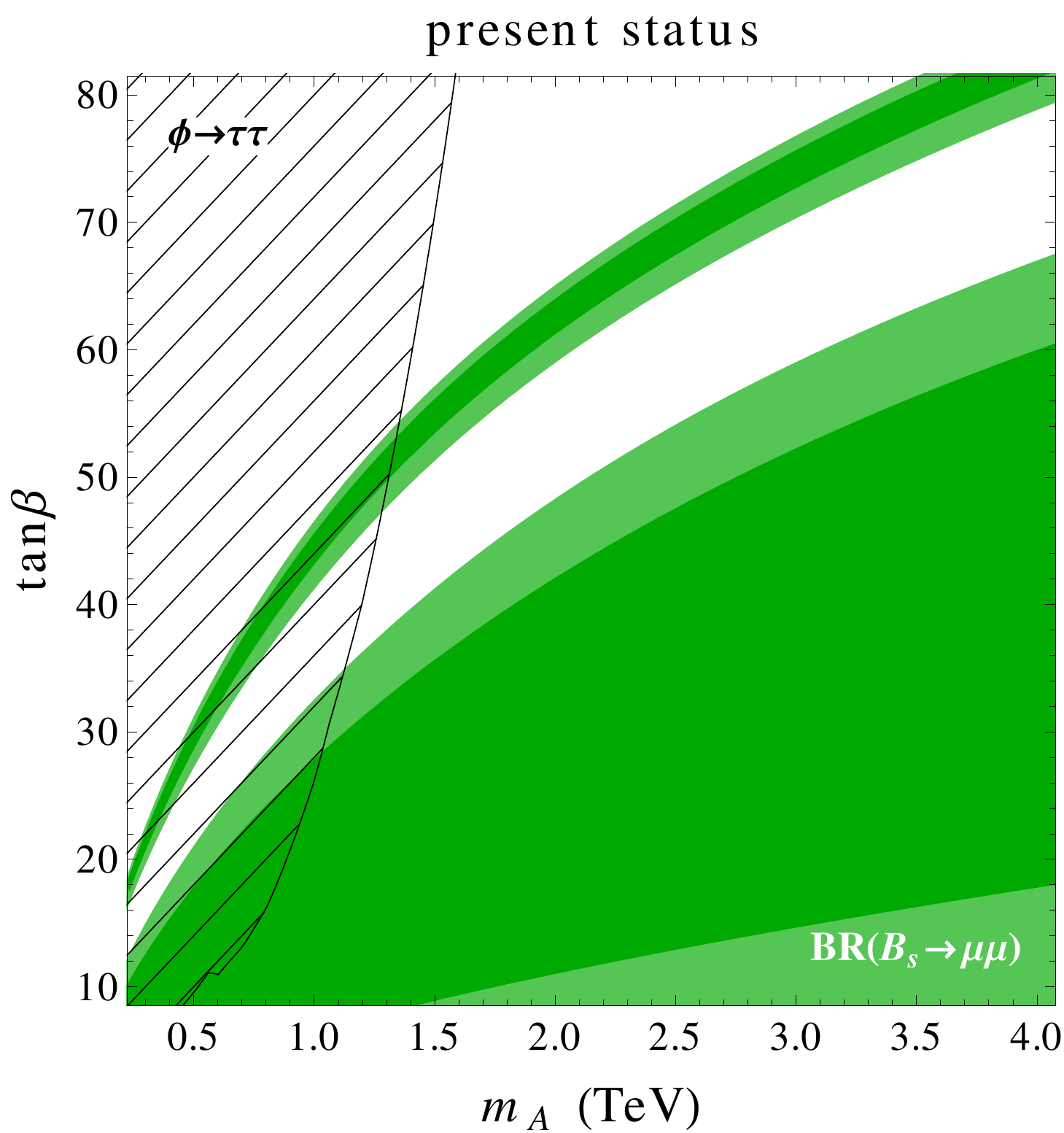}
\caption{Current constraints in the $m_A$ - $\tan\beta$ plane in the MSSM scenario discussed in the text. The dark and light green shaded regions are {\it allowed} by the $\overline{\text{BR}}(B_s \to \mu^+\mu^-)$ measurements at the $1\sigma$ and $2\sigma$ level. The black hatched region is {\it excluded} by direct searches for $\tau^+\tau^-$ resonances. Throughout the plot the light Higgs mass is $m_h = 125$~GeV.}
\label{fig:susy}
\end{figure}

The sensitivity of the current branching ratio measurements to MSSM parameter space is illustrated in Fig.~\ref{fig:susy}. The dark and light green regions correspond to the regions where $\overline{\text{BR}}(B_s \to \mu^+\mu^-)$ is compatible with the measurements at the $1\sigma$ and $2\sigma$ level. The white region is excluded by $\overline{\text{BR}}(B_s \to \mu^+\mu^-)$ by more than $2\sigma$.
We observe two distinct regions of parameter space. As expected, there is (i) a broad region for small $\tan\beta$ and large $m_A$ corresponding to a NP amplitude $A \ll 1$, and (ii) a thin stripe for larger values of $\tan\beta$ where $A \simeq 1$ that also agrees well with the measured branching ratio.

\begin{figure}[tb]
\centering
\includegraphics[width=0.46\textwidth]{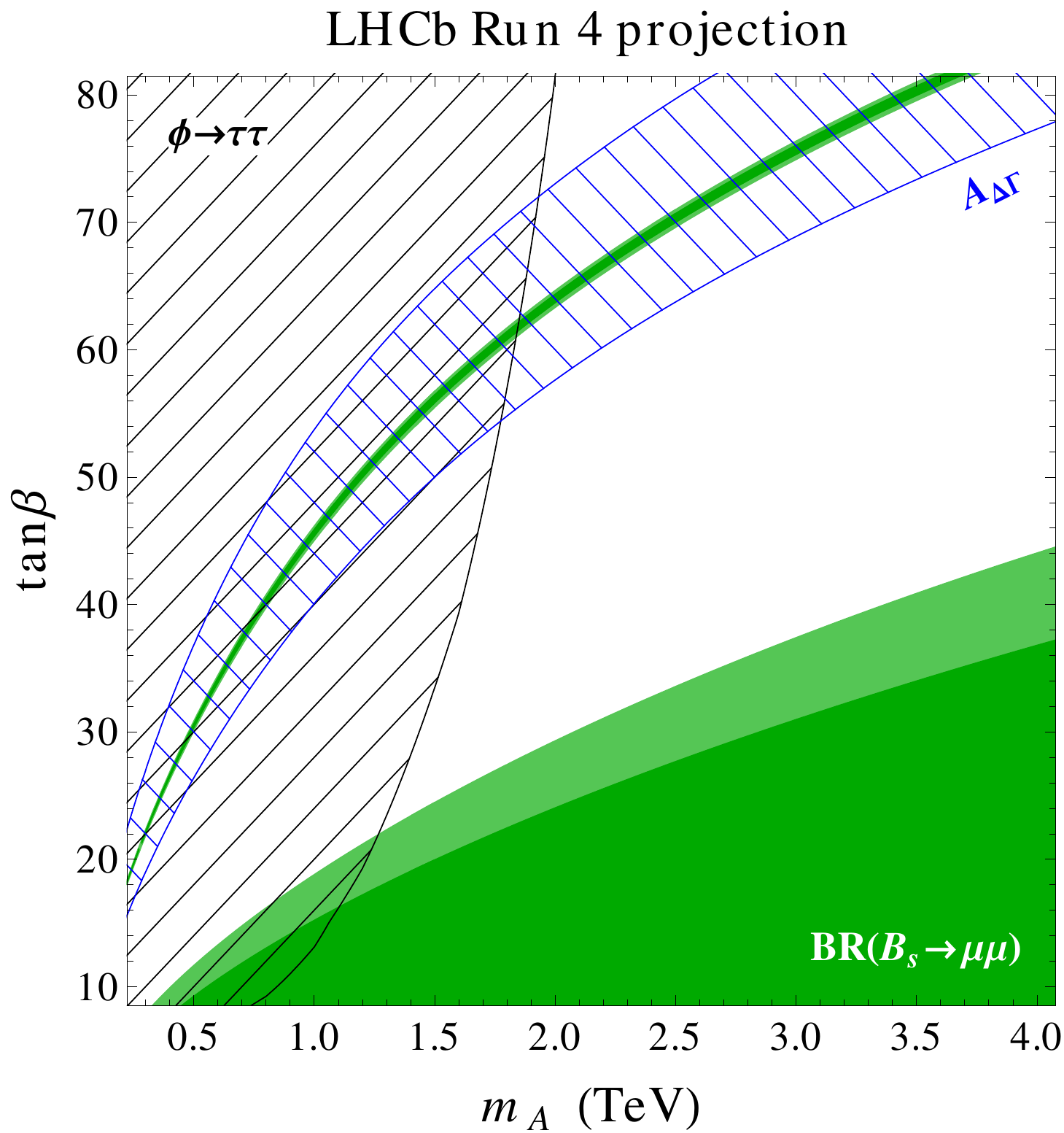} ~~~
\includegraphics[width=0.46\textwidth]{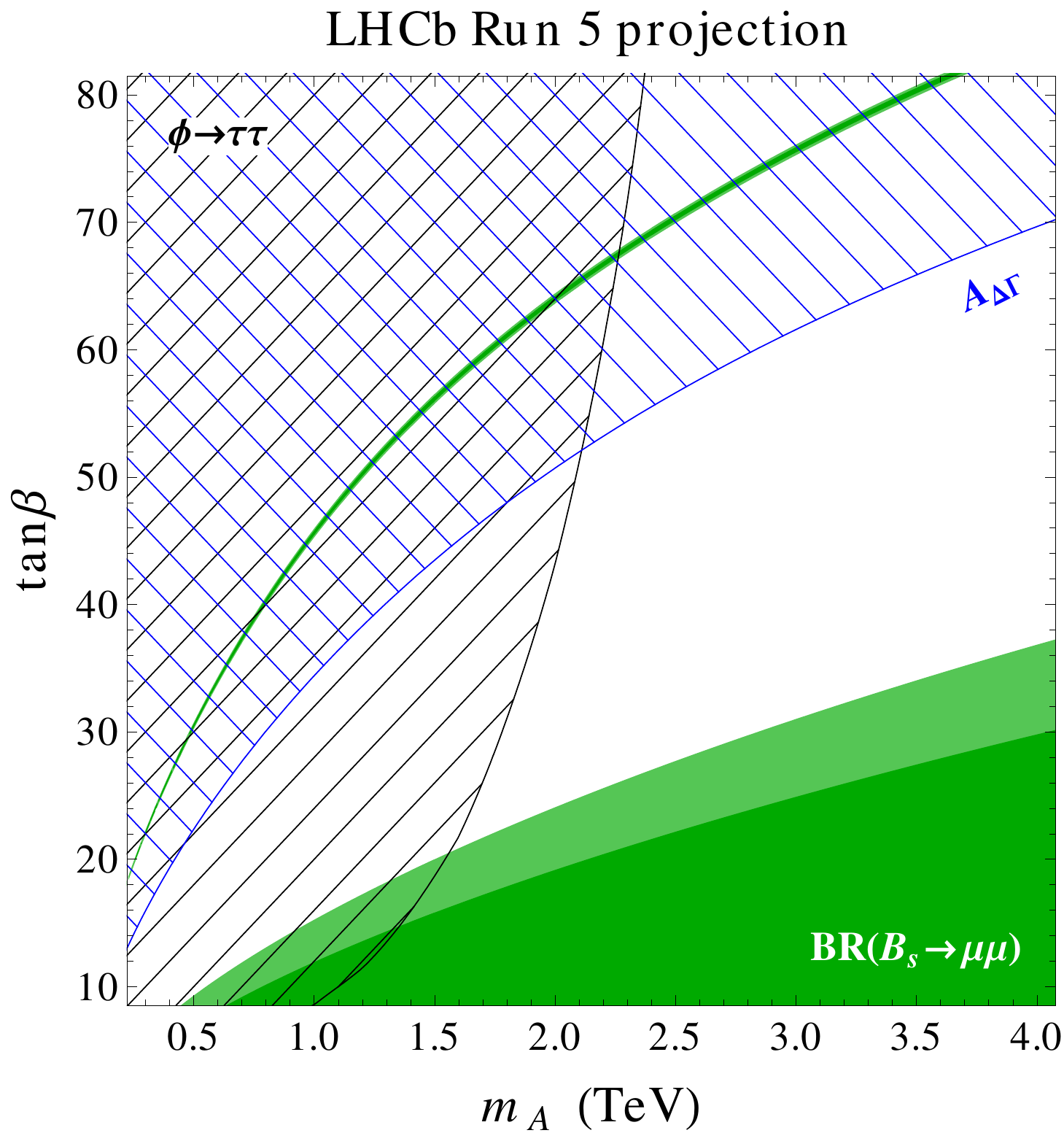}
\caption{Expected sensitivities in the $m_A$ - $\tan\beta$ plane in the MSSM scenario discussed in the text. Left: integrated luminosities of 50~fb$^{-1}$ at LHCb and 300~fb$^{-1}$ at CMS and ATLAS. Right: integrated luminosities of 300~fb$^{-1}$ at LHCb and 3000~fb$^{-1}$ at CMS and ATLAS. The dark and light green shaded regions will be {\it allowed} by the expected $\overline{\text{BR}}(B_s \to \mu^+\mu^-)$ sensitivity at the $1\sigma$ and $2\sigma$ level, assuming the SM rate. The black hatched region could be {\it excluded} by direct searches for $\tau^+\tau^-$ resonances assuming no non-standard signal. The blue hatched region can be covered by measurements of the mass-eigenstate rate asymmetry $A_{\Delta \Gamma}$. In both  plots the light Higgs mass is $m_h = 125$~GeV.}
\label{fig:susy2}
\end{figure}

In the plots of Fig.~\ref{fig:susy2} we show the $m_A$ - $\tan\beta$ plane in the two future scenarios discussed above. While the size of the $A \ll 1$ region and the $A \simeq 1$ stripe is shrinking with more precise data, the branching ratio measurement alone cannot exclude the $A \simeq 1$ scenario that corresponds to a sizable new physics contribution.
The sensitivity of future measurements of the mass-eigenstate rate asymmetry $A_{\Delta \Gamma}$ is also shown in the plots. The blue hatched regions correspond to $A_{\Delta \Gamma} < -0.6$ (left plot) and $A_{\Delta \Gamma} < 0.4$ (right plot).
We can clearly see that that future measurements of $A_{\Delta \Gamma}$ can cover unconstrained parameter space and fully probe the $A \simeq 1$ region.

Finally, we discuss the complementarity of the $B_s \to \mu^+\mu^-$ observables and direct searches for the heavy Higgs bosons.
The main production modes of heavy neutral Higgs bosons $H$ and $A$ in the MSSM are either gluon fusion or, at large $\tan\beta$, production in association with $b$ quarks. In the parameter regions that we are mainly interested in, namely multi-TeV Higgs bosons and large $\tan\beta$, we find that the production in association with $b$ quarks is by far dominant.

The corresponding production cross section can be easily obtained by rescaling known SM results
\begin{equation}
 \sigma_{b \bar b}(H/A) = \frac{t_\beta^2}{(1 + \epsilon_b t_\beta)^2} \times \sigma_{b \bar b}(H/A)_\text{SM} ~,
\end{equation}
where $\sigma_{b \bar b}(H/A)_\text{SM}$ is the production cross section of $H/A$ with SM like couplings to $b$ quarks,  and the $\epsilon_b$ parameter was already given above.
The $\sigma_{b \bar b}(H/A)_\text{SM}$ cross section depends only on the mass of the neutral Higgs bosons and we compute it at NNLO using the public code \verb|bbh@nnlo|~\cite{Harlander:2003ai}.

Concerning the heavy Higgs decays, we note that multi-TeV Higgs bosons are sufficiently close to the decoupling limit, such that we can neglect decays of the scalar $H$ to massive gauge bosons $WW$ and $ZZ$ and decays of the pseudoscalar into $A \to Zh$. We also neglect decays into two light Higgs bosons $H \to hh$ (which is $\tan\beta$ suppressed) and $A \to hh$ (which is non-zero only in the presence of CP violation).
In our setup, all other SUSY particles are sufficiently heavy such that ``exotic'' decays for example into neutralinos $H \to \chi^0 \chi^0$, or staus $H \to \tilde \tau^+ \tilde \tau^-$ are not kinematically open.
In this case, the main decay modes are $H/A \to t\bar t, b \bar b, \tau^+\tau^-$.
For low $\tan\beta$, the decays to tops dominate. For large $\tan\beta$ one has roughly $90\%$ branching ratio to $b\bar b$ and $10\%$ branching ratio to $\tau^+\tau^-$.
We approximate the total decay width as sum of the top, bottom and tau decay widths.
The relevant expressions are
\begin{eqnarray}
 \Gamma(H/A \to t\bar t) &=& \frac{1}{t_\beta^2} \times \Gamma(H/A \to t\bar t)_\text{SM} ~, \\
 \Gamma(H/A \to b\bar b) &=& \frac{t_\beta^2}{(1 + \epsilon_b t_\beta)^2} \times \Gamma(H/A \to b\bar b)_\text{SM} ~, \\
 \Gamma(H/A \to \tau^+ \tau^-) &=& \frac{t_\beta^2}{(1 + \epsilon_\tau t_\beta)^2} \times \Gamma(H/A \to \tau^+\tau^-)_\text{SM} ~.
\end{eqnarray}
In the decay to $t \bar t$, we do not include higher-order non-holomorphic corrections. Those become relevant only for large $\tan\beta$, where the $t \bar t$ width itself is negligibly small.
For the decay widths with SM like couplings $\Gamma(H/A \to f\bar f)_\text{SM}$, we take into account NLO QCD corrections in the heavy Higgs mass limit from~\cite{Gorishnii:1990zu}. We find
\begin{eqnarray}
 \frac{\Gamma(H \to b\bar b)_\text{SM}}{\Gamma(H \to \tau^+\tau^-)_\text{SM}} &\simeq& \frac{\Gamma(A \to b\bar b)_\text{SM}}{\Gamma(A \to \tau^+\tau^-)_\text{SM}} \simeq \frac{3 m_b^2}{m_\tau^2} \left( 1 + \frac{17}{3} \frac{\alpha_s}{\pi} \right) ~, \\
 \frac{\Gamma(H \to t\bar t)_\text{SM}}{\Gamma(H \to b\bar b)_\text{SM}} &\simeq&  \frac{\Gamma(A \to t\bar t)_\text{SM}}{\Gamma(A \to b\bar b)_\text{SM}}\left( 1 - \frac{4 m_t^2}{m_H^2} \right) \simeq \frac{m_t^2}{m_b^2} \left( 1 - \frac{4 m_t^2}{m_H^2} \right)^\frac{3}{2} ~.
\end{eqnarray}
The fermion masses and $\alpha_s$ in the above expressions are $\overline{\text{MS}}$ values at the scale of the heavy Higgs boson mass.

The most sensitive searches for MSSM Higgs bosons look for the $\tau^+\tau^-$ final state.
Constraints are given on the cross section times $\tau^+\tau^-$ branching ratio separately for gluon fusion production  and production in association with b quarks.
To obtain constraints on the heavy Higgs parameter space we add up the signal cross sections from $H$ and $A$, $\sigma_{b \bar b}(H) \times \text{BR}(H \to \tau^+\tau^-) + \sigma_{b \bar b}(A) \times \text{BR}(A \to \tau^+\tau^-)$ and compare to the latest 13~TeV bounds from~\cite{CMS:2016rjp,ATLAS:2016fpj}. For each value of the heavy Higgs mass, we use the stronger of the CMS and ATLAS bounds.
The current constraints are shown in Fig.~\ref{fig:susy} as black hatched regions.
We observe that for Higgs masses below $\sim 1.1$~TeV, the constraints from the direct searches are stronger than the indirect ones from $B_s \to \mu^+\mu^-$. For larger Higgs masses, however, $B_s \to \mu^+\mu^-$ covers unconstrained parameter space.

For the future scenarios we expect that the sensitivity of the direct searches will extend to considerably higher masses. We extrapolate the current expected sensitivities by scaling them with the square root of appropriate luminosity ratios.
In the left plot of Fig.~\ref{fig:susy2} we show the expected reach of the direct searches with 300~fb$^{-1}$ of data, in the right plot the expected reach corresponds to 3000~fb$^{-1}$ of data.

The plots clearly demonstrate the complementarity of the direct searches, the branching ratio measurement, and the measurement of $A_{\Delta \Gamma}$. All three probes are required to maximise the coverage of the $m_A$ - $\tan\beta$ parameter space.

\section{\BsmumuHeading\ in Leptoquark Models}\label{sec:lq}

Models with leptoquarks (LQs), i.e.\ scalar or vector states
coupling to a quark-lepton current,
have received lots of attention in recent years
as they could explain various anomalies in $B$ physics data,
including angular observables in $B\to K^*\mu^+\mu^-$
and hints for lepton flavour universality violation in $B \to K\mu^+\mu^-$
and $B \to D^{(*)} \tau \nu$
(see e.g.\ \cite{Hiller:2014yaa, Bauer:2015knc, Fajfer:2015ycq, Barbieri:2015yvd, Becirevic:2015asa, Becirevic:2016yqi, Barbieri:2016las}).
Assuming SM gauge invariance, the possible quantum numbers for LQs are restricted to ten possible representations (of which five with spin~0 and five with
spin~1) \cite{Buchmuller:1986zs}.
In this work we are interested in scenarios that generate the scalar operators
contributing to $B_s\to\mu^+\mu^-$.
At tree level, this is only the case for two representations,
\begin{itemize}
\item the vector LQ $U_1$ transforming as
$({\mathbf{3}}, \mathbf{1})_{\frac{2}{3}}$ under $G_\text{SM}=SU(3)_c\times SU(2)_L\times U(1)_Y$  and
\item the vector LQ $V_2$ transforming as
$(\bar{\mathbf{3}}, \mathbf{2})_{\frac{5}{6}}$.
\end{itemize}
Due to the strong enhancement of $B_s\to\mu^+\mu^-$ in the presence of scalar
operators, we have also investigated whether any of the other eight LQ
representations could give rise to the scalar $bs\mu\mu$ operators at
the one-loop level. We found that the only non-zero contribution comes from
loop diagrams including Higgs exchange, making these contributions completely negligible.
Thus we will restrict our discussion to the two vector LQs $U_1$ and
$V_2$ and to tree-level effects in the following.
Interestingly, the $U_1$ LQ is also able to explain the
$b \to s \ell \ell$ anomalies \cite{Barbieri:2015yvd,Barbieri:2016las}.

For the two considered scenarios, $U_1$ and $V_2$, the interaction Lagrangians are given
in the $G_\text{SM}$-invariant interaction basis by~\cite{Buchmuller:1986zs}
\begin{subequations}
\begin{align}
 \mathcal{L}_{U_1} &= \hat\lambda_\text{L}^{i j} \, \left( \bar{Q}_\text{L}^i \, \gamma_\mu \, L_\text{L}^j \right) U_1^\mu +
                      \hat\lambda_\text{R}^{i j} \, \left( \bar{d}_\text{R}^i \, \gamma_\mu \, e_\text{R}^j \right) U_1^\mu + \text{h.c.},\\
 \mathcal{L}_{V_2} &= \hat\lambda_\text{R}^{i j} \, \bar{d}_\text{R}^{\mathsf{c} \, i}  \, \gamma_\mu \, \left( L_\text{L}^j \cdot \epsilon \cdot V_2^\mu  \right) +
                      \hat\lambda_\text{L}^{i j} \, \left( \bar{Q}_\text{L}^{\mathsf{c} \, i} \cdot \epsilon \cdot V_2^\mu \right) \, \gamma_\mu \, e_\text{R}^j +
                      \hat\lambda_{qq}^{i j} \, \left( \bar{Q}_\text{L}^{\mathsf{c} \, i} \cdot V_2^{* \,\mu} \right) \gamma_\mu \, u_\text{R}^j + \text{h.c.},
\end{align} \label{eq:LQLint}
\end{subequations}
where $i,j$ are generational indices.
After EWSB the quark fields have to be rotated into the mass basis in order to get to the effective operators in the weak Hamiltonian.
This can be done by defining LQ couplings in the mass basis that one obtains by contracting the above gauge basis couplings with the quark rotation matrices.
We will denote the mass basis couplings without the hat, e.g. ${\lambda}_\text{L}^{b \mu} = \left[ V_{d \text{L}}^\dagger \right]^{3 i} \hat\lambda_\text{L}^{i 2}$.

\begin{table}[tbp]
\begin{center}
\renewcommand{\arraystretch}{1.4}
\renewcommand\tabcolsep{12pt}
\begin{tabular}{ccc}
\hline\hline
 & $U_1$ & $V_2$ \\
 \hline
 $C_9^\text{NP}$     & $-\frac{1}{2} \mathcal{N} {\lambda}_L^{s\mu} {\lambda}_L^{b\mu \, *}$            &  $\frac{1}{2} \mathcal{N} {\lambda}_L^{b\mu} {\lambda}_L^{s\mu \, *}$            \\
 $C_9'$              & $-\frac{1}{2} \mathcal{N} {\lambda}_R^{s\mu} {\lambda}_R^{b\mu \, *}$            &  $\frac{1}{2} \mathcal{N} {\lambda}_R^{b\mu} {\lambda}_R^{s\mu \, *}$            \\
 $C_{10}^\text{NP}$  &  $\frac{1}{2} \mathcal{N} {\lambda}_L^{s\mu} {\lambda}_L^{b\mu \, *}$            &  $\frac{1}{2} \mathcal{N} {\lambda}_L^{b\mu} {\lambda}_L^{s\mu \, *}$            \\
 $C_{10}'$           & $-\frac{1}{2} \mathcal{N} {\lambda}_R^{s\mu} {\lambda}_R^{b\mu \, *}$            & $-\frac{1}{2} \mathcal{N} {\lambda}_R^{b\mu} {\lambda}_R^{s\mu \, *}$            \\
 $C_S=-C_P$          &              $\mathcal{N} {\lambda}_L^{s\mu} {\lambda}_R^{b\mu \, *} \, m_b^{-1}$&              $\mathcal{N} {\lambda}_R^{b\mu} {\lambda}_L^{s\mu \, *} \, m_b^{-1}$\\
 $C_S'=C_P'$         &              $\mathcal{N} {\lambda}_R^{s\mu} {\lambda}_L^{b\mu \, *} \, m_b^{-1}$&              $\mathcal{N} {\lambda}_L^{b\mu} {\lambda}_R^{s\mu \, *} \, m_b^{-1}$ \\
\hline\hline
\end{tabular}
\end{center}
\caption{The $bs\mu\mu$ Wilson coefficients in the two LQ models in terms of mass-basis
couplings. The superscript ``NP'' denotes the new physics contribution.
The normalization factor $\mathcal N$ is defined in~\eqref{eq:LQN}.}
\label{tab:lq}
\end{table}

In terms of these couplings we find the Wilson coefficients\footnote{We note that we disagree with \cite{Dorsner:2016wpm} on a factor of $\frac{1}{2}$ for the scalar Wilson coefficients for $V_2$. Our result, however, is consistent with \cite{Hiller:2016kry}.} in Tab.~\ref{tab:lq},
where the normalization coefficient is given as
\begin{equation}
 \mathcal{N} = -\frac{2 \pi }{\alpha V_{ts}^* V_{tb}} \left( \frac{v}{m_\text{LQ}} \right)^2.
 \label{eq:LQN}
\end{equation}
where $v=246\,\text{GeV}$ denotes the SM Higgs VEV.
Since these couplings are invariant under $G_\text{SM}$, the tree-level
coefficients satisfy the SMEFT relations \eqref{eq:GSMinvariance}.
We observe that both scenarios further satisfy the relation
\begin{equation} \label{eq:LQ_WCrelations}
 m_b^2 \, C_S \, C_S'
 =  -4 \, C_{10}^\text{NP} \, C_{10}'
 =  -4 \, C_{9}^\text{NP} \, C_{9}'
\end{equation}
while
$U_1$ fulfills $C_{9}^\text{NP}=C_{10}^\text{NP}$ and $C_{9}'=-C_{10}'$
and
$V_2$ fulfills $C_{9}^\text{NP}=-C_{10}^\text{NP}$ and $C_{9}'=C_{10}'$.

\subsection{Direct searches for vector leptoquarks}

Leptoquarks are color charged and therefore can be pair produced by the strong interaction at the LHC.
Assuming that all other couplings of the LQs to SM particles are sufficiently small compared to the strong gauge coupling, pair production through QCD is the dominant production mode.
While for scalar LQs, $S$, $SU(3)$ gauge invariance completely fixes their couplings to gluons,
for vector LQs, $V_\mu$, there is more freedom
and the production cross section becomes more model dependent.
Only if the vector LQs are gauge bosons of an extended gauge group, gauge invariance does completely determine the couplings to gluons.
In the most general case the interactions read~\cite{Blumlein:1996qp,Rizzo:1996ry}
\begin{equation} \label{eq:VLQ_int}
 \mathcal L_\text{int} = -\frac{1}{2} V_{\mu\nu}^\dagger V^{\mu\nu} - i g_s (1-\kappa) V_\mu^\dagger G^{\mu\nu} V_\nu - i g_s \tilde \kappa V_\mu^\dagger \tilde G^{\mu\nu} V_\nu ~,
\end{equation}
with $V_{\mu\nu} = D_\mu V_\nu - D_\nu V_\mu$ and $G^{\mu\nu}$, $\tilde G^{\mu\nu}$ are the gluon field strength and the dual field strength, respectively.
For vector LQs that are gauge bosons the anomalous couplings vanish, $\kappa = \tilde \kappa = 0$.
The term proportional to $\tilde \kappa$ breaks CP.
We assume CP conservation and neglect this interaction.
The anomalous coupling $\kappa$ is a free parameter.
Therefore, we will consider two scenarios:
\begin{enumerate}
 \item the gauge boson case with $\kappa = 0$,
 \item the maximally conservative choice where $\kappa$ is fixed in such a way that it minimizes the production cross section.
\end{enumerate}
The full expressions for the partonic cross sections for vector LQs can be found e.g.\ in~\cite{Blumlein:1996qp}.
In our numerical analysis we obtain the LQ production cross section at the LHC by convoluting the partonic cross section with CT14 NNLO parton distribution functions \cite{Dulat:2015mca} using \texttt{LHAPDF6} \cite{Buckley:2014ana}.

The decay modes of the LQ are model-dependent as they are determined by the couplings of the LQ to SM fields.
We consider the minimal case, where we only allow for couplings that are relevant for $b s \mu\mu$ transitions.
Including also other couplings would decrease the branching ratios and therefore weaken the constraints from direct searches.

In the case of the $U_1$ LQ we consider the decays $U_1 \rightarrow s \mu$ and $U_1 \rightarrow b \mu$.
Invariance under $\text{SU}(2)_\text{L}$ also implies the existence of the LQ coupling to up-type quarks and neutrinos (see eq. (\ref{eq:LQLint})).
These couplings are related to the $b\mu$, $s\mu$ couplings via the CKM matrix
\begin{equation}
 {\lambda}_\text{L}^{u \nu_\mu} = V_{us} \, {\lambda}_\text{L}^{s\mu} \,\, + \,\, V_{ub} \, {\lambda}_\text{L}^{b\mu},~~~
 {\lambda}_\text{L}^{c \nu_\mu} = V_{cs} \, {\lambda}_\text{L}^{s\mu} \,\, + \,\, V_{cb} \, {\lambda}_\text{L}^{b\mu},~~~
 {\lambda}_\text{L}^{t \nu_\mu} = V_{ts} \, {\lambda}_\text{L}^{s\mu} \,\, + \,\, V_{tb} \, {\lambda}_\text{L}^{b\mu}.
\end{equation}
Therefore the relevant decay channels are
\begin{equation}
 U_1 \rightarrow \mu j, \quad U_1 \rightarrow \nu j, \quad U_1 \rightarrow \nu t,
\end{equation}
where $j$ denotes a jet coming from any quark but the top.

The $V_2$ LQ is a $SU(2)$ doublet and consists of a charge $4/3$ and $1/3$ state.
Following a line of reasoning similar to the $U_1$ case, we find the relevant decay modes to be
\begin{equation}
 V_2^\frac{1}{3} \rightarrow \mu j, \quad V_2^\frac{4}{3} \rightarrow \mu j, \quad V_2^\frac{1}{3} \rightarrow \nu j.
\end{equation}
In principle, there can be a mass difference between the two $V_2$ components after EWSB, such that the heavier state can decay into the lighter emitting a $W$ boson.
Any mass splitting has to be proportional to the Higgs vev and therefore should be at most of the order of
\begin{equation}
 \Delta m_{V_2} \sim \frac{v^2}{2 m_{V_2}} \sim 30~\text{GeV} \times \left( \frac{1 ~\text{TeV}}{m_{V_2}} \right) ~.
\end{equation}
This implies that the $W$ is off-shell for realistic LQ masses and we can neglect the branching ratio for such a process.
For the $V_2$ LQ also a coupling to two quarks is allowed (rf. eq. (\ref{eq:LQLint})), such that a decay into two jets could be important.
Such an interaction would violate baryon number and, therefore, induce proton decay which is highly constrained.
In the following, we will simply assume baryon number conservation and neglect this type of coupling.

Most experimental LQ searches target pair-produced LQs that decay into a quark and a lepton.
The relevant signature is $jj\ell\ell$ (or $jj\ell + \slashed{E}_T$ if one LQ decays into a neutrino).
A common assumption in the experimental searches is that the LQ couples only to one generation of SM particles.
However, the searches for first and second generation LQs usually do not apply a $b$-veto such that also the $b \mu$ final states relevant for this work are covered.
A last caveat is that most searches implicitly assume scalar LQs.
In the case of vector LQs, the angular distributions will be different, leading to slightly different acceptances. This effect has been found to be small in~\cite{Khachatryan:2015vaa} and we will neglect it in the following.
In our analysis we include the LQ searches from ATLAS \cite{ATLAS:2014sga,Aaboud:2016qeg} and CMS \cite{CMS:2016qhm} using $8 \, \text{TeV}$ and $13 \, \text{TeV}$ data.

\subsection{Leptoquark effects in \BsmumuHeading}

The LQ scenario $U_1$ can explain present anomalies in semi-leptonic $b\to s\mu\mu$ transitions \cite{Barbieri:2015yvd} if $C_9 = -C_{10} \approx -0.5$ and $C_9' = C_{10}' \approx 0$ \cite{Altmannshofer:2014rta,Descotes-Genon:2015uva,Hurth:2016fbr}.
In the following we will focus on this case.
We fix the left-handed couplings such that $C_9$ and $C_{10}$ take the above values and keep the right-handed couplings as free parameters.
By (\ref{eq:LQ_WCrelations}) we are then forced to set either the unprimed or primed scalar operator to zero.
We decide to consider the following benchmark scenario:\footnote{Note that the requirement of $C_9 = -C_{10} \approx -0.5$ does not uniquely determine the values of the left-handed couplings. Only the product of ${\lambda}_\text{L}^{s\mu}$ and ${\lambda}_\text{L}^{b\mu}$ is fixed, but not their ratio. We assume that both are of the same order.}
\begin{equation}U_1: \qquad\qquad
        {\lambda}_\text{L}^{s\mu} = {\lambda}_\text{L}^{b\mu} = + \frac{1}{\sqrt{\mathcal{N}(m_\text{LQ})}}, \quad
	{\lambda}_\text{R}^{s\mu} = 0, \quad
	{\lambda}_\text{R}^{b\mu} \text{~~free,}\qquad\qquad\qquad \label{eq:LQU1}
\end{equation}
which results in
$C_9 = -C_{10} = -0.5$, $C_9'=C_{10}'=0$, $C_S=-C_P$ (free), $C_S'=C_P'=0$.

The LQ $V_2$ predicts $C_9^\text{NP} = +C_{10}^\text{NP}$ and therefore cannot explain the $b \to s \mu \mu$ anomalies.
We assume degenerate masses for the two components with charges $\frac{1}{4}$ and $\frac{4}{3}$, and consider the following scenario:
  \begin{equation}V_2: \qquad\qquad
   {\lambda}_\text{L}^{s \mu} = {\lambda}_\text{L}^{b \mu} = \sqrt{\frac{2 \times 0.1}{\mathcal{N}(m_\text{LQ})}}, \quad
   {\lambda}_\text{R}^{s \mu} = 0, \quad
   {\lambda}_\text{R}^{b \mu}  \text{~~free.} \qquad\qquad\qquad \label{eq:LQV2}
  \end{equation}
This setup corresponds to a rather small value of $C_9 = +C_{10} = 0.1$, vanishing primed Wilson coefficients, and free $C_S = -C_P$.

\begin{figure}
  \includegraphics[keepaspectratio=true,width=0.5\textwidth]{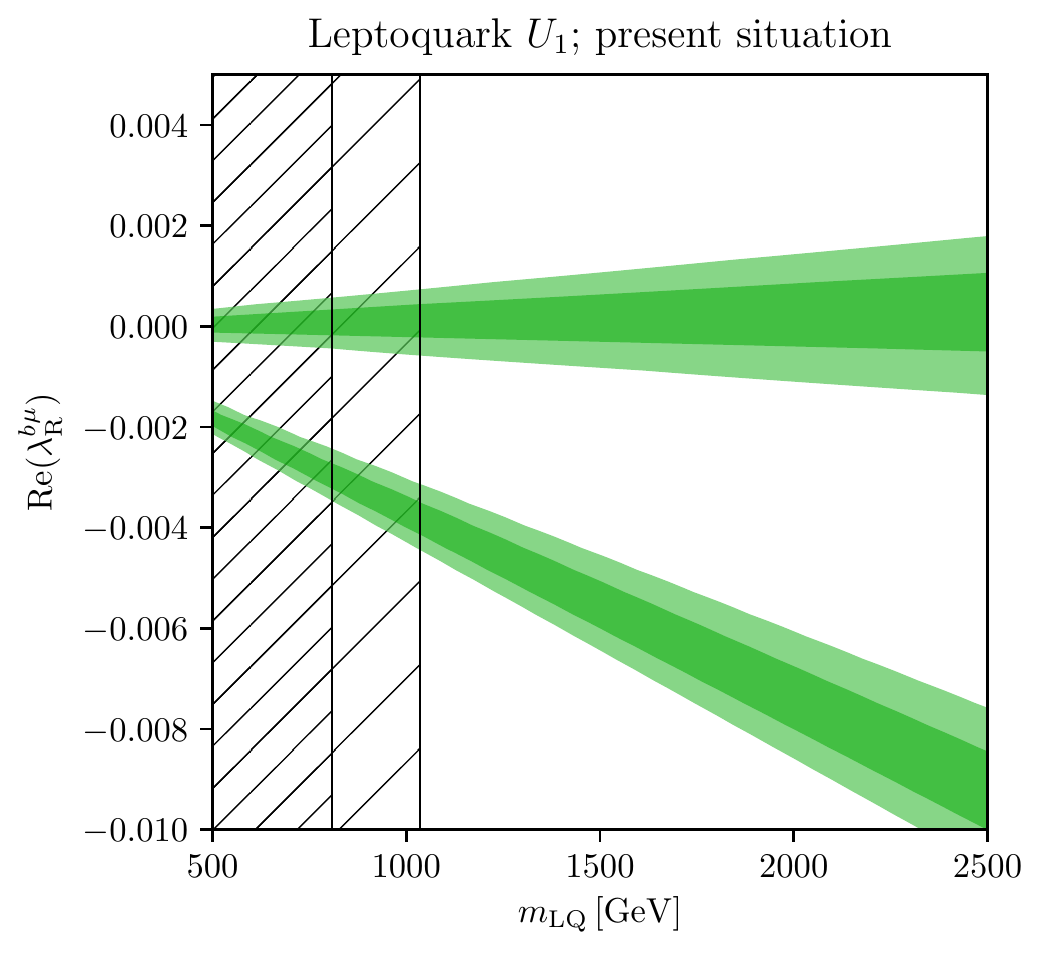}%
 \includegraphics[keepaspectratio=true,width=0.5\textwidth]{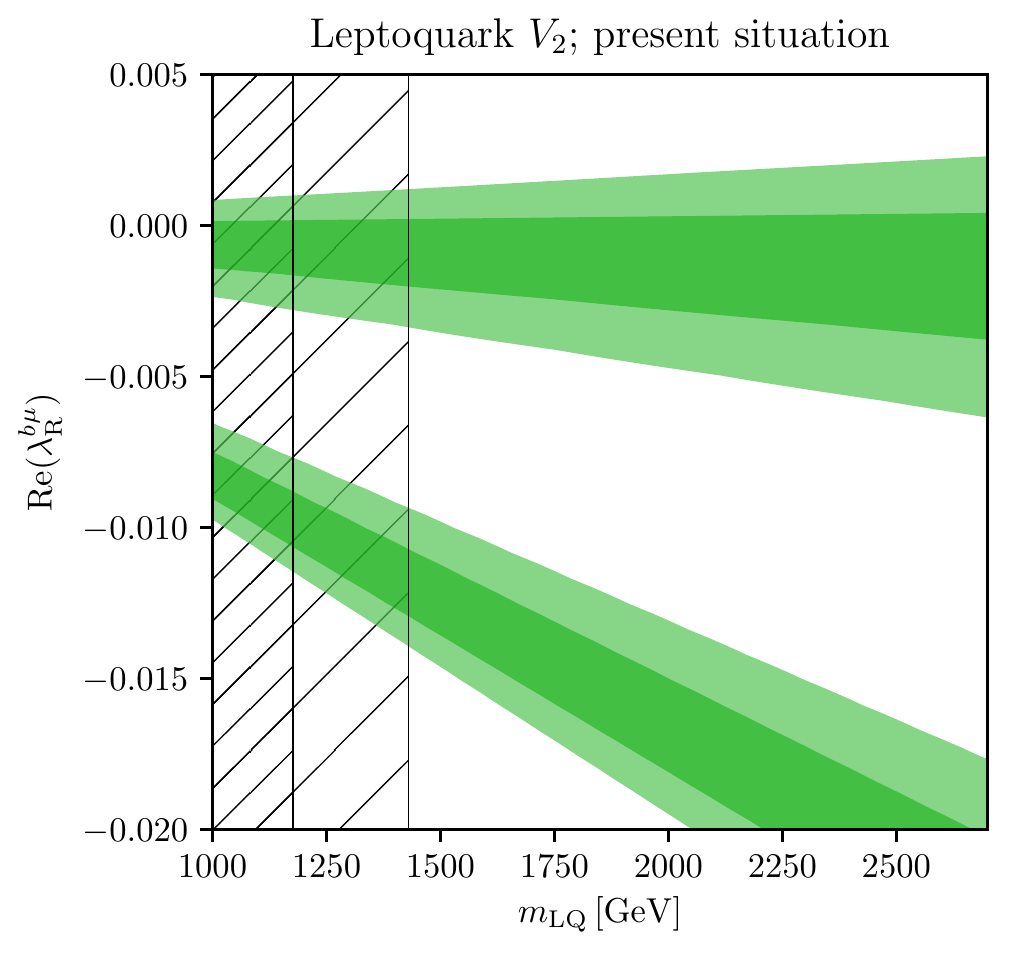}
 \caption{The currently allowed parameter regions in the mass vs. coupling plane for the LQs $U_1$ (\emph{left}) and $V_2$ (\emph{right}) in the scenarios (\ref{eq:LQU1}) and (\ref{eq:LQV2}).
          Inside the dark and light green bands, the present value of the  experimental branching ratio (\ref{eq:ourav}) is reproduced at 1 and $2\sigma$, respectively.
          The black //-hatched regions show the exclusions from present direct searches. The more densely hatched region corresponds to minimal LQ production, while the more coarsely hatched region is for YM-like production.}
 \label{fig:LQpresent}
\end{figure}
In Fig. \ref{fig:LQpresent} we show the presently allowed parameter regions for both scenarios in the mass-coupling plane.
In the parameter space still allowed by experimental searches one finds two solutions that reproduce the current measurement of the branching ratio.
One is SM-like while the other corresponds to large NP effects.

\begin{figure}
 \centering
 \includegraphics[keepaspectratio=true,width=0.5\textwidth]{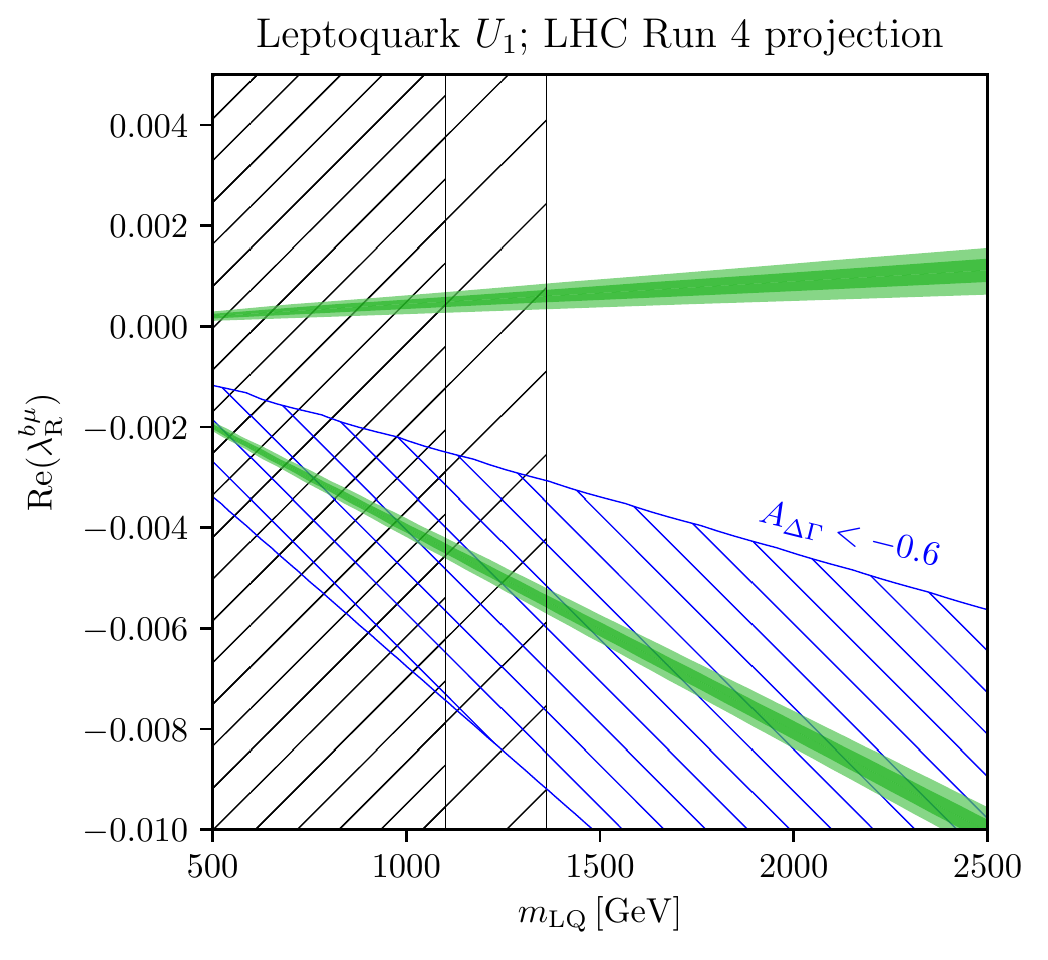}%
 \includegraphics[keepaspectratio=true,width=0.5\textwidth]{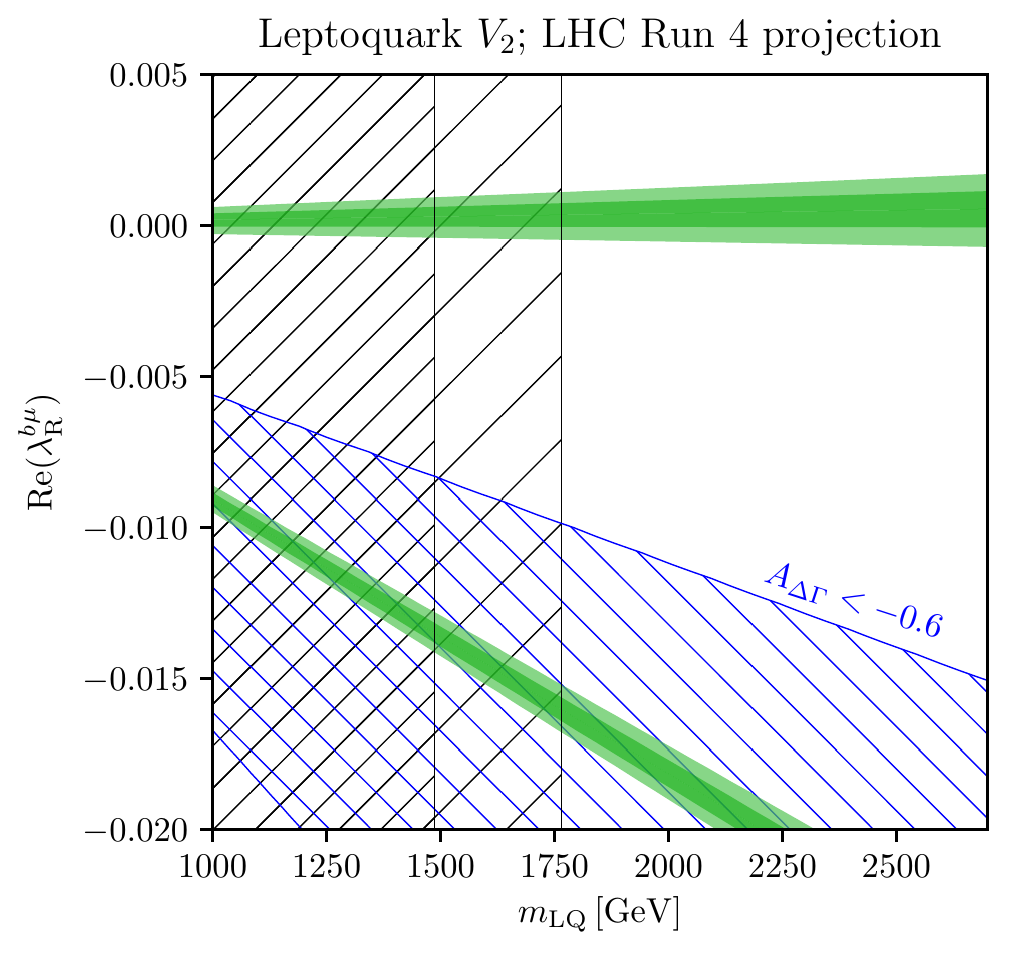} \\%
 \includegraphics[keepaspectratio=true,width=0.5\textwidth]{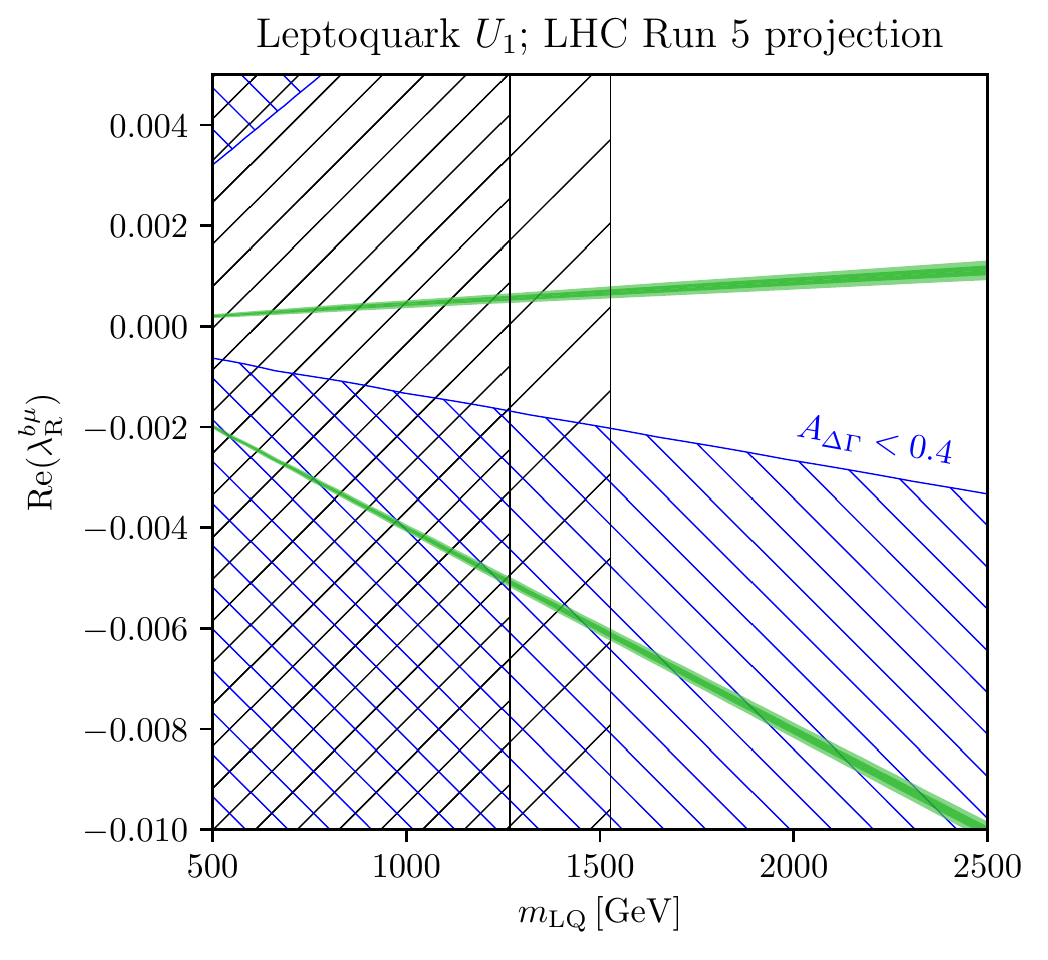}%
 \includegraphics[keepaspectratio=true,width=0.5\textwidth]{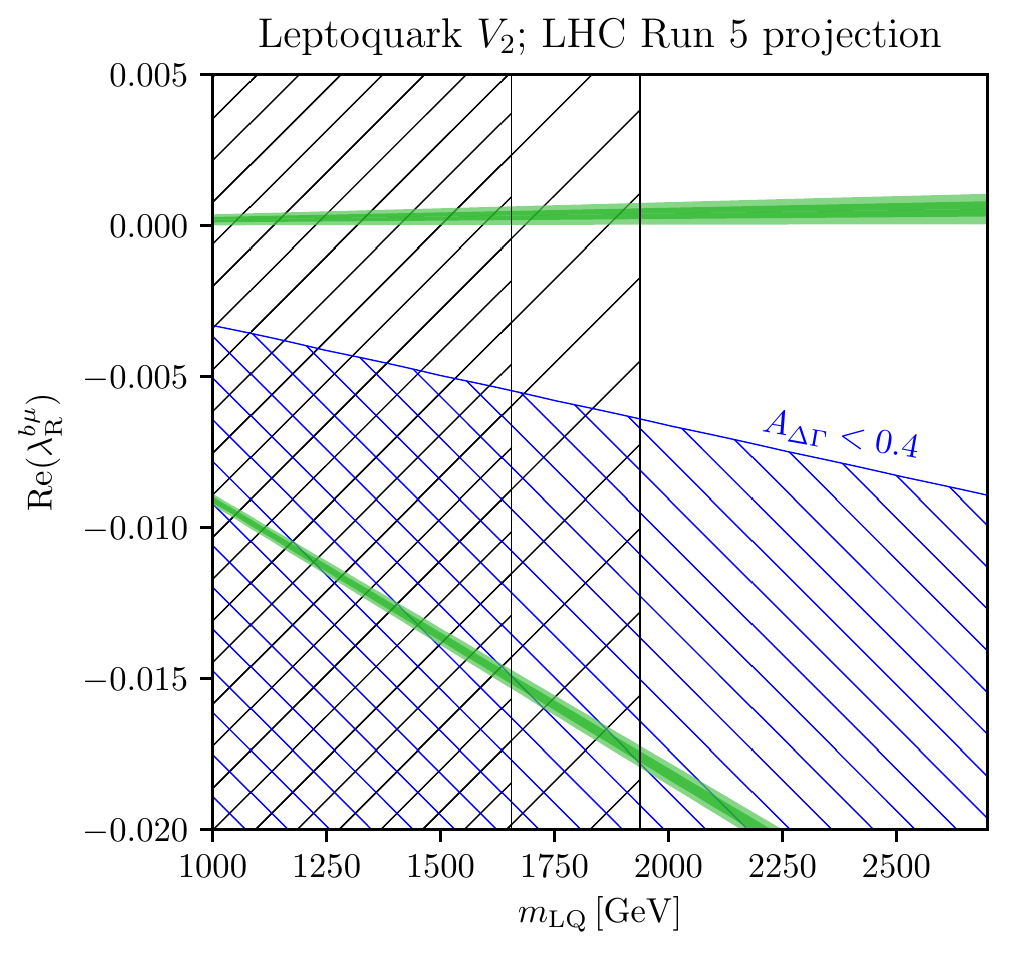}
 \caption{Future constraints in the mass vs. coupling planes for the LQs $U_1$ (\emph{left}) and $V_2$ (\emph{right}) in the scenarios (\ref{eq:LQU1}) and (\ref{eq:LQV2}).
	  The first row is for the ``Run 4'' scenario while the second row marks ``Run 5''.
	  The green $1\sigma$- and $2\sigma$-regions correspond to the anticipated future experimental sensitivity of the branching ratio, assuming the SM central value (\ref{eq:BRSM}).
          The black //-hatched regions show the extrapolated exclusions from direct searches. The more densely hatched region corresponds to minimal LQ production, while the more coarsely hatched region is for YM-like production.
          The blue \textbackslash\textbackslash-hatched region would be excluded at the $2\sigma$ level by a measurement of $\ADG$ with SM-like central value.}
 \label{fig:PlotsLQProjections}
\end{figure}

Projecting into the future, a measurement of $\ADG$ will be able to disentangle this situation.
We find that the NP solution gives rise to a large negative $\ADG$, such that already an estimation of its sign can rule out this scenario.
In Fig. \ref{fig:PlotsLQProjections} we present future projections for the mass vs. coupling plane.
We consider the Run 4 and 5 of the LHC.
We estimate the direct constraints by rescaling the present exclusion limits by the square root of the luminosity ratios, $\sqrt{\mathcal{L}_\text{today} / \mathcal{L}_\text{future}}$, where $ \mathcal{L}_\text{future} = 300 \, \text{fb}^{-1}$, $3000 \, \text{fb}^{-1}$ for Run 4 and Run 5, respectively.
In high mass ranges, for which presently there are no direct constraints, we conservatively extrapolate the current exclusion limits as constants.
For the branching ratio we assume the current SM value (\ref{eq:BRSM}) with uncertainties (\ref{eq:UncertaintyProjections}).
These projections clearly show the power of an $\ADG$ measurement in eliminating a degenerate solution as well as the general impact of $B_s \to \mu^+\mu^-$ on the LQ parameter space.

The LQ representation $V_2$ also contributes to $B \to K^{(*)} \nu \nu$ \cite{Buras:2014fpa}.
Using the notation of that reference we find that $B_s \to \mu^+ \mu^-$ is correlated to the $b \to s \nu\nu$ transition via $m_\frac{1}{3}^2 C_R =  m_\frac{4}{3}^2 C_9' = - m_\frac{4}{3}^2  C_{10}'$.
As NP contributions to $C_9' = -C_{10}'$ are strongly constrained, one cannot expect large contributions to $B \to K^{(*)} \nu \nu$.
Taking a benchmark value of $C_9' = -C_{10}' = 0.33$ (which corresponds to the maximally allowed value at the $2\sigma$-level found in \cite{Altmannshofer:2014rta}) and assuming degenerate masses for the two LQ components, we only find a modification of about $3 \%$ relative to the SM expectation.
This, however, is only true if the LQ solely couples to muons.
If also couplings to other lepton generations are allowed, then the effects can be much larger
\cite{Buras:2014fpa}.

In summary, we have shown that a measurement of $\ADG$ can efficiently reduce degeneracies in the LQ parameter space that cannot be resolved by a measurement of the branching ratio alone.
Interestingly, even the two-fold ambiguity remaining in the presence of
simultaneous real NP effects in $C_S$ and $C_S'$
as shown in Fig. \ref{fig:modelindependent:OneWC}
might be resolved in the LQ scenarios due to the relation \eqref{eq:LQ_WCrelations}
that predicts simultaneous effects in $C_9$ and $C_9'$ as well as $C_{10}$ and
$C_{10}'$ for this solution, which could be tested in semi-leptonic $B$ decays
such as $B\to K^*\mu^+\mu^-$. Whether their sensitivity will be strong enough
to really resolve this ambiguity is an interesting question but beyond the scope
of our present study.

\section{Conclusions}\label{sec:conclusions}

In this work we discussed the new physics sensitivity of present and future measurements of the $B_s\to\mu^+\mu^-$ decay.

The $B_s\to\mu^+\mu^-$ decay is a rare flavour-changing neutral current process and widely recognized as one of the most important indirect probes of new physics at the LHC.
The present experimental measurements of the branching ratio are compatible with the SM prediction and have uncertainties at the level of 20\%.
With data taken in future runs of the LHC, the experimental precision can improve significantly, potentially reaching $\sim 5\%$ with 50~fb$^{-1}$ in Run 4 at LHCb, and $2\%$ with 300~fb$^{-1}$ in Run 5 at LHCb, respectively. The SM prediction is expected to reach similar precision within one decade.
In addition to high precision measurements of the branching ratio, the large statistics that will become available in future LHC runs will also allow the experiments to measure additional observables in the $B_s\to\mu^+\mu^-$ decay.
In particular, a measurement of the mass-eigenstate rate asymmetry $A_{\Delta \Gamma}$ with a precision of $80\%$ in Run 4 and $30\%$ in Run 5 should become possible.

We analysed the interplay of the branching ratio and the mass-eigenstate rate asymmetry in constraining new physics both model-independently and in the context of new physics models, namely the minimal supersymmetric standard model and two leptoquark models.
Our analysis shows that future measurements of the mass-eigenstate rate asymmetry would allow to disentangle new physics scenarios with scalar interactions, that would be indistinguishable based on measurements of the branching ratio alone.

Under mild assumptions, scalar contributions to the $B_s\to\mu^+\mu^-$ decay can be model-independently described by two complex parameters. Fits to branching ratio data alone leave a large degeneracy in parameter space. Assuming CP conservation, the degeneracy can be broken by $A_{\Delta \Gamma}$ up to a two-fold ambiguity.
We provide all the tools
necessary to reproduce the plots from our model-independent
analysis in Sec.~\ref{sec:modelindependent} with
\texttt{flavio} in a public repository \cite{plotrepo}.

In the context of the MSSM, we use the measurements of BR$(B_s\to\mu^+\mu^-)$ to obtain constraints in the $m_A$-$\tan\beta$ plane. We find that the branching ratio alone is not sensitive to a region of parameter space with a sizable SUSY contribution to $B_s\to\mu^+\mu^-$ predicting opposite sign but similar magnitude for the transition amplitude. Future measurements of $A_{\Delta \Gamma}$ can cover this region. We also highlight the complementarity with direct searches for heavy Higgs bosons. While the direct searches put the strongest constraints at low Higgs masses $\lesssim 1$~TeV, the indirect constraints from $B_s\to\mu^+\mu^-$ extend the coverage far into the multi-TeV region.

We make similar observations also in the considered leptoquark models. There are two leptoquark representations
that can mediate scalar operators
contributing to $B_s\to\mu^+\mu^-$. In both models, and similarly to the MSSM, there are regions of parameter space with sizable contributions to the $B_s\to\mu^+\mu^-$ amplitude that lead to a SM-like branching ratio. Measurements of $A_{\Delta \Gamma}$ can cover much of these regions, although ambiguities remain in the most general case. Direct searches for leptoquarks exclude most regions of parameter space for leptoquark masses of around 1~TeV and below. The $B_s\to\mu^+\mu^-$ observables are able to indirectly probe much heavier leptoquarks.

Our work explicitly demonstrates the complementarity of the $B_s\to\mu^+\mu^-$ branching ratio and mass-eigenstate rate asymmetry in probing new physics and stresses the relevance of the large statistics that can be expected from future high luminosity runs at the LHC.

\section*{Acknowledgements}

WA acknowledges discussions with Stefania Gori and financial support by the University of Cincinnati.
Special thanks go to Peter Stangl for discussions and helping to digitize experimental plots using \texttt{svg2data} \cite{peter_stangl_2017_292635}.
CN would like to thank the University of Cincinnati for the hospitality during large parts of this work.
DS would like to thank Flavio Archilli, Marc-Oliver Bettler, Patrick Koppenburg, and Sandro Palestini for useful discussions
on the LHCb and ATLAS measurements.
The work of CN and DS was supported by the DFG cluster of excellence ``Origin and Structure of the Universe''.

\bibliographystyle{JHEP}
\bibliography{bibliography}

\end{document}